\documentstyle[11pt,amssymb,epsf]{article}

\makeatletter
\@addtoreset{equation}{section}
\makeatother

\def\ben{\begin{equation}}
\def\een{\end{equation}}

\let\a=\alpha    
    
  \let\n=\nu   
 \let\t=\tau    
         \let\L=\Lambda
     
\let\C=\Chi

\def\nn{\nonumber} \def\bd{\begin{document}} \def\ed{\end{document}}
\def\ds{\documentstyle} \let\fr=\frac \let\bl=\bigl \let\br=\bigr
\let\Br=\Bigr \let\Bl=\Bigl
\let\bm=\bibitem
\let\na=\nabla
\let\pa=\partial \let\ov=\overline
\newcommand{\be}{\begin{equation}}
\newcommand{\ee}{\end{equation}}
\def\ba{\begin{array}}
\def\ea{\end{array}}
\def\ft#1#2{{\textstyle{{\scriptstyle #1}\over {\scriptstyle #2}}}}
\def\fft#1#2{{#1 \over #2}}
\def\del{\partial}
\def\vp{\varphi}
\def\sst#1{{\scriptscriptstyle #1}}
\def\oneone{\rlap 1\mkern4mu{\rm l}}
\def\td{\tilde}
\def\wtd{\widetilde}
\def\ie{\rm i.e.\ }
\def\dalemb#1#2{{\vbox{\hrule height .#2pt
        \hbox{\vrule width.#2pt height#1pt \kern#1pt
                \vrule width.#2pt}
        \hrule height.#2pt}}}
\def\square{\mathord{\dalemb{6.8}{7}\hbox{\hskip1pt}}}
\newcommand{\ho}[1]{$\, ^{#1}$}
\newcommand{\hoch}[1]{$\, ^{#1}$}
\newcommand{\bea}{\begin{eqnarray}}
\newcommand{\eea}{\end{eqnarray}}
\newcommand{\ra}{\rightarrow}
\newcommand{\lra}{\longrightarrow}
\newcommand{\Lra}{\Leftrightarrow}
\newcommand{\ap}{\alpha^\prime}
\newcommand{\bp}{\tilde \beta^\prime}
\newcommand{\tr}{{\rm tr} }
\newcommand{\Tr}{{\rm Tr} }
\def\0{{\sst{(0)}}}
\def\1{{\sst{(1)}}}
\def\2{{\sst{(2)}}}
\def\3{{\sst{(3)}}}
\def\4{{\sst{(4)}}}
\def\5{{\sst{(5)}}}
\def\6{{\sst{(6)}}}
\def\7{{\sst{(7)}}}
\def\8{{\sst{(8)}}}
\def\n{{\sst{(n)}}}
\def\cA{{{\cal A}}}
\def\cF{{{\cal F}}}
\def\tV{\widetilde V}
\def\tW{\widetilde W}
\def\tH{\widetilde H}
\def\tE{\widetilde E}
\def\tF{\widetilde F}
\def\tA{\widetilde A}
\def\im{{{\rm i}}}
\def\tY{{{\wtd Y}}}
\def\ep{{\epsilon}}
\def\vep{{\varepsilon}}
\def\R{\rlap{\rm I}\mkern3mu{\rm R}}
\def\bD{{{\bar D}}}

\def\R{\rlap{\rm I}\mkern3mu{\rm R}}
\def\bD{{{\bar D}}}
\def\R{{{\Bbb R}}}
\def\C{{{\Bbb C}}}
\def\H{{{\Bbb H}}}
\def\CP{{{\Bbb C}{\Bbb P}}}
\def\RP{{{\Bbb R}{\Bbb P}}}
\def\Z{{{\Bbb Z}}}
\def\bA{{{\Bbb A}}}
\def\bB{{{\Bbb B}}}
\def\bC{{{\Bbb C}}}
\def\bR{{{\Bbb R}}}
\def\bD{{{\Bbb D}}}
\def\bE{{{\Bbb E}}}
\def\bZ{{{\Bbb Z}}}
\def\cD{{{\cal D}}}
\def\Re{{{\frak{Re}}}}
\def\Im{{{\frak{Im}}}}
\def\cosec{{\,\hbox{cosec}\,}}
\def\Gm{{\Gamma_{\!\! -}}}
\def\Gp{{\Gamma_{\!\! +}}}

\def\stan{{standard }}
\def\nonstan{{supernumerary }}

\def\cosech{{\hbox{cosech}}}

\def\etcyc{{\hbox{and cyclic}}}
\def\btheta{{\bar\theta}}

\newcommand{\tamphys}{\it Center for Theoretical Physics,
Texas A\&M University, College Station, TX 77843, USA}
\newcommand{\umich}{\it Michigan Center for Theoretical Physics,
University of Michigan\\ Ann Arbor, MI 48109, USA}
\newcommand{\upenn}{\it Department of Physics and Astronomy,\\
University of Pennsylvania, Philadelphia,  PA 19104, USA}
\newcommand{\SISSA}{\it  SISSA-ISAS and INFN, Sezione di Trieste\\
Via Beirut 2-4, I-34013, Trieste, Italy}

\newcommand{\mitchell}{\it George P. \& Cynthia W. 
Mitchell Institute for Fundamental Physics,\\
Texas A\&M University, College Station, TX 77843-4242, USA}

\newcommand{\newton}{\it Isaac Newton Institute for Mathematical Sciences,\\
0 Clarkson Road,  University of Cambridge,
Cambridge CB3 0EH, UK}

\newcommand{\ihp}{\it Institut Henri Poincar\'e\\
  11 rue Pierre et Marie Curie, F 75231 Paris Cedex 05}

\newcommand{\damtp}{\it DAMTP, Centre for Mathematical Sciences,
 Cambridge University,\\  Wilberforce Road, Cambridge CB3 OWA, UK}
\newcommand{\itp}{\it Institute for Theoretical Physics, University of
California\\ Santa Barbara, CA 93106, USA}

\newcommand{\auth}{
M. Cveti\v c\hoch{\dagger}, G.W. Gibbons\hoch{\star}, 
H. L\"u\hoch{\ddagger} and C.N. Pope\hoch{\ddagger}}

\begin{document}
\begin{flushright}
\hfill {
DAMTP-2003-49\ \ \
MIFP-03-12\ \ \
UPR/0852-T
}\\
\hfill{
June\ \ 2003,\ \ \ \ \ \bf hep-th/0306043}
\end{flushright} 

\begin{center}  

{\large {\bf Consistent Group and Coset Reductions of the Bosonic String}}   

\vspace{12pt}

\auth

\vspace{10pt}
  {\hoch{\dagger} {\it School of Natural
Sciences,\\ Institute for Advanced Studies,
 Princeton, NJ 08540, USA}}

\vspace{7pt}
 {\hoch{\star}\damtp}

\vspace{7pt}
{\hoch{\ddagger}\mitchell}

\vspace{14pt}

\underline{ABSTRACT}
\end{center}  

    Dimensional reductions of pure Einstein gravity on cosets other
than tori are inconsistent.  The inclusion of specific additional
scalar and $p$-form matter can change the situation.  For example, a
$D$-dimensional Einstein-Maxwell-dilaton system, with a specific
dilaton coupling, is known to admit a consistent reduction on $S^2=
SU(2)/U(1)$, of a sort first envisaged by Pauli.  We provide a new
understanding, by showing how an $S^3=SU(2)$ group-manifold reduction
of $(D+1)$-dimensional Einstein gravity, of a type first indicated by
DeWitt, can be broken into in two steps; a Kaluza-type reduction on
$U(1)$ followed by a Pauli-type coset reduction on $S^2$.  More
generally, we show that any $D$-dimensional theory that itself arises
as a Kaluza $U(1)$ reduction from $(D+1)$ dimensions admits a
consistent Pauli reduction on any coset of the form $G/U(1)$.
Extensions to the case $G/H$ are given.  Pauli coset reductions of the
bosonic string on $G= (G\times G)/G$ are believed to be consistent,
and a consistency proof exists for $S^3=SO(4)/SO(3)$.  We examine
these reductions, and arguments for consistency, in detail.  The
structures of the theories obtained instead by DeWitt-type
group-manifold reductions of the bosonic string are also studied,
allowing us to make contact with previous such work in which only
singlet scalars are retained.  Consistent truncations with two singlet
scalars are possible.  Intriguingly, despite the fact that these are
not supersymmetric models, if the group manifold has dimension 3 or 25
they admit a superpotential formulation, and hence first-order
equations yielding domain-wall solutions.

\pagebreak
\setcounter{page}{1}

\tableofcontents
\addtocontents{toc}{\protect\setcounter{tocdepth}{2}}
\newpage

\section{Introduction}

     Dimensional reductions, sometimes referred to as Kaluza-Klein
reductions, have had a long and convoluted history.  Shortly after the
inception of general relativity, Kaluza in 1919 pointed out that by
considering a circle reduction of five-dimensional Einstein gravity,
one obtains a unification of gravity with Maxwell theory in four
dimensions, at the cost of introducing an extra scalar field
\cite{kaluza1}.  In subsequent developments this scalar field was
viewed as somewhat of an embarrassment, and it was usually arbitrarily
(and incorrectly) set to a constant value.  This was done, for
example, in the work of Klein \cite{klein1} (he did, however, make the
important observation that the charges are quantised).  In fact it was
not until the much later work of Jordan (1947) and Thiry (1948) that
it was fully appreciated that if one imposes the five-dimensional
Einstein equations, subject solely to the ``cylindrical condition''
then if the Maxwell field is non-vanishing it is dynamically
inconsistent to impose the condition that the scalar is constant
\cite{jordan,thiry}.

    The next development came with the work of Pauli in 1953, who
attempted in an unpublished work to obtain $SU(2)$ Yang-Mills fields
from a reduction scheme in which a six-dimensional spacetime is taken
to be the product of a 2-sphere (with its round metric) and a
four-dimensional spacetime \cite{pauli}.\footnote{An examination of
Klein's 1938 paper on Yang-Mills gauge invariance reveals that, as far
as gravity is concerned, he definitely did not have in mind a
reduction either on a group manifold or a coset space.  His
higher-dimensional spacetime was only five-dimensional, and his metric
appears to be non-commutative \cite{klein2,gross,oraf}.  Thus although
Klein made pioneering contributions to non-Abelian gauge theories, the
gauge bosons did not originate from higher dimensional symmetries.}
Nowadays, generalisations of this model are in widespread use, and are
often referred to as ``coset reductions.''  In this paper we shall
call them ``Pauli Reductions.''  Pauli realised that there are
considerable problems in obtaining the Yang-Mills field equations from
the six-dimensional Einstein equations in a fully consistent dynamical
way.  He seems to have realised, for example, that there was no
justification for substituting his ansatz into the six-dimensional
action functional.  This point has often been overlooked in subsequent
developments.

   It appears that it was DeWitt in 1963 who was the first to obtain
the Yang-Mills equations in $n$ dimensions from a reduction scheme in
which one assumes that $(n+q)$-dimensional spacetime is invariant
under the action of a compact semi-simple Lie group $G$ of dimension
$q$ \cite{dewitt}.  Such reductions are frequently referred to as
group-manifold reductions; in this paper we shall call them ``DeWitt
Reductions.''  In the exercise in \cite{dewitt}, it was assumed that
the metric on the orbit space of the group $G$ was the bi-invariant
metric, but DeWitt indicated that a fully consistent reduction would
involve taking a metric which was merely left-invariant.  This results
in $\ft12 q(q+1)$ scalar fields propagating in the lower-dimensional
spacetime.  Indeed these are the analogues of the scalar field that
was often omitted in the older work on five-dimensional theories, and
again one can easily see that their omission is inconsistent with the
higher-dimensional vacuum Einstein equations.  These fields, which
parameterise the shape of the ``internal'' group manifold, are often
referred to as ``scalar moduli.'' It is a curious fact that if
one takes the higher-dimensional theory to be not pure Einstein
gravity but rather the low-energy effective action of the bosonic
string, then a DeWitt reduction exists in which the scalar moduli can
be consistently omitted \cite{het1}.

     In 1968 Kerner, apparently unaware of DeWitt's
work, showed explicitly that by substituting the ``restricted DeWitt
ansatz'' (\ie without the scalar moduli) into the higher-dimensional
Einstein action one obtains a lower-dimensional action for gravity
plus Yang-Mills fields \cite{kerner}.  This calculation is of course
correct in so far as it goes, but ignores the ``consistency issue,''
which is whether solutions of the equations of motion of the
lower-dimensional action do actually provide solutions of the
higher-dimensional Einstein equations.  In fact in this case they do
not.  Later, the theory of consistent DeWitt reductions (\ie including the
scalar moduli) was developed by Cho and Freund (1975).  Further work by
Scherk and Schwarz (1979) emphasised the importance of requiring that
the group $G$ be unimodular (meaning that the structure 
constants satisfy $f^\a{}_{\a\beta}=0$) \cite{chofre,schsch}.  The
condition of unimodularity is automatically satisfied for compact
Lie groups, which in practice is where our interest lies.

    Somewhat earlier than the work of Cho and Freund and of Scherk and
Schwarz, it was realized by Hawking (1969) \cite{hawking} working on
homogeneous Bianchi cosmology, i.e. DeWitt reductions to one
spacetime dimension, that substitution of the ansatz into the Einstein
action will not always give the correct field equations. Roughly
speaking, the point is that to obtain the the field equations, an
integration by parts is required.  If the internal space is closed,
i.e. compact without boundary, this presents no obvious
problem. However if the internal space is non-compact, since all
variations are, according to the ansatz, $G$-invariant, one cannot
merely assume, as one ordinarily does, that they vanish outside a
compact set or ``at infinity.'' Following a period of confusion,
the situation was clarified by Sneddon (1976) \cite{sneddon}, who
showed in detail that if the group $G$ is unimodular (known as class A
in the Bianchi classification), then no problem arises and the
procedure works.  However, if the group is non-unimodular (known as
class B in the Bianchi classification) then problems can and do arise
and incorrect equations of motion are obtained. A simple definition of
unimodular is that the adjoint representation of the group has unit
determinant, or infinitesimally that $f^\a{}_{\a\beta}=0$.  An equivalent
definition is that any left-invariant measure on $G$ is also
right-invariant.  Thus all top-dimensional forms on $G$ are
proportional up a factor which does not depend upon the coordinates on
$G$. Yet another way of stating the condition is that the generators
of right translations, that is the left-invariant vector fields, have
vanishing divergence.  A semi-simple group (compact or not) is
necessarily unimodular and a compact group is necessarily unimodular
and so perhaps because they limited their attention to that case, Cho
and Freund encountered no difficulty.  Scherk and Schwarz by contrast
considered more general groups and drew attention, apparently unaware
of the work on Bianchi cosmology, to the need for unimodularity.  In
the context of dimensional reduction one is almost always interested
in compact internal spaces, whether group manifolds or not, and so
integration by parts will then be a valid manoeuvre.

    The situation for Pauli reductions, \ie coset reductions, is
considerably more subtle than that for DeWitt reductions, and to date 
there exist very few known
examples of such reductions that are consistent.  Of course, if we
assumed that the higher-dimensional spacetime were invariant under the
action of a Lie group $G$ acting on a coset $K=G/H$, then
consistency would be expected.\footnote{In this paper, all cosets will
be what is customarily (but not completely universally) called {\sl 
right cosets}, \ie we quotient $G$ by the equivalence relation
$g_1\equiv g_2$ if and only if there is an $h\in H$ such that $g_1 =
h\, g_2$.  Thus right cosets admit global right actions of the group
$G$, but in general no left action.}  However, the aim of the exercise is to
obtain fields in the lower-dimensional spacetime that include all of
the gauge bosons of the isometry group $G$.  In order to do this,
it is necessary to make an ansatz for the higher-dimensional
metric and other fields that is {\sl not} invariant under the action
of $G$.  On the other hand, we are obviously not interested in going to the
opposite extreme, by retaining {\sl all} the fields in a generalised
Fourier expansion, which would certainly guarantee consistency, but
would have little point since it would merely provide a clumsy
description of an intrinsically higher-dimensional situation.  Thus an
important ingredient when considering a coset reduction could be said
to be that one requires a reduction to a {\sl finite} set of
lower-dimensional fields that includes {\sl all} the gauge bosons
associated with the isometry group of the coset.  Having established
the desiderata for the ansatz, the statement of consistency is as
follows: Substitution of the reduction ansatz into the
higher-dimensional equations of motion must yield a set of equations
of motion expressed entirely in terms of the lower-dimensional fields.
In other words, if $x^\mu$ are coordinates on the lower-dimensional
spacetime, and $y^m$ are coordinates on the internal space, then 
substituting the reduction ansatz into the higher-dimensional
equations of motion should result in a system of equations in which
all $y^m$ dependence has cancelled.
These lower-dimensional equations may or may not be derivable from a
lower-dimensional action principle, which itself may or may not be obtained 
by substitution of the ansatz into the higher-dimensional action.  
It should be emphasised, however, that the test of consistency
is determined solely by the equations of motion, and is quite independent
of any considerations of action principles.

    At present there is no known algorithmic prescription for
obtaining consistent Pauli reduction ans\"atze.  A systematic
understanding of the conditions under which such reductions are
possible is still lacking.  In the few examples where consistent Pauli
reductions are known, there has hitherto been little conceptual
understanding of why they work beyond the bald statement that detailed
calculation shows that they do.

   The first two candidates yielding consistent Pauli reductions in
fact arose in the compactification of eleven-dimensional supergravity.
The first was the $S^7$ reduction to four-dimensional $SO(8)$ gauged
$N=8$ supergravity, for which a proof of consistency was presented in
\cite{dwinic}; and the second was the $S^4$ reduction to
$SO(5)$-gauged $N=4$ supergravity in seven dimensions, for which the
consistency proof was presented in \cite{nasvamvan}.  It is widely
believed that a consistent Pauli reduction of type IIB supergravity on
$S^5$ also exists, which would yield $SO(6)$-gauged $N=8$ supergravity
in five dimensions.  A consistent $S^5$ Pauli reduction of a
truncation of type IIB supergravity to its $SL(2,\R)$-singlet sector,
yielding a five-dimensional theory with all the gauge bosons of
$SO(6)$, has been constructed in \cite{s5red}.  Various explicit
consistent reductions yielding subsets of the maximally-supersymmetric
cases have also been constructed (see, for example,
\cite{con1,con2,con3,con4}).

    The consistent Pauli reductions mentioned above might lead one to
conclude that supersymmetry was an essential ingredient in the
consistency of the Pauli reduction.  However, various further examples
of consistent Pauli reductions exist which lie entirely outside the
framework of supergravity.  They do still, however, require that one
consider a higher-dimensional theory that goes beyond pure Einstein
gravity.  Two such theories were considered in \cite{spherered}, one
being the low-energy effective action for the $D$-dimensional bosonic
string, and the other being a specific $D$-dimensional theory of
Einstein gravity coupled to a Maxwell field and a dilaton.  It was
shown that a consistent Pauli reduction of the $D$-dimensional bosonic
string on $S^3$ or $S^{D-3}$ is always possible, and that a
consistent Pauli reduction of the $D$-dimensional
Einstein-Maxwell-dilaton theory on $S^2$ is always possible
\cite{spherered}.  It should be emphasised that the $S^3$ reduction of
the bosonic string is of Pauli type, and not merely of DeWitt type;
it yields the full set of $SO(4)$ gauge fields.  This latter
example lends support to an old conjecture in \cite{het2}, 
asserting that a consistent Pauli reduction of the bosonic string on the
group manifold $G$ should always be possible;
again, keeping the full set of gauge bosons of the $G\times G$
isometry group of the bi-invariant metric on $G$.

\section{The Bosonic String}

A  bosonic string moving in background
fields  $G_{MN}, B_{MN}, \Phi$ in $D$ dimensions 
 may be regarded as a generalized 2-dimensional non-linear sigma model.
The demand  that the beta functions vanish gives rise to
the equations of motion for  the background fields and these
equations of motion may in fact be derived from a stationary
 action principle. To lowest order in the inverse string tension
$1 \over 2 \pi \alpha ^\prime$, one has \cite{callan}
\bea
16 \pi ^2 \beta ^{\Phi} &=& {D-26 \over 3 \alpha ^\prime} +\Big [4
(\nabla \Phi)^2 -4 \nabla ^2 \Phi -R + {1 \over 12 } H^2 \Big ]\,,\nn\\
\beta ^G_{MN}&=& R_{MN} - {1 \over 4} H_M \,^{PQ} H_{NPQ} + s \nabla _M
\nabla _N \Phi\,,\label{betafn}\\
\beta ^B_{MN}&=& \nabla _P H^P\, _{MN} -2 H^P _{MN} \nabla _P \Phi\,,\nn
\eea
where the 3-form $H_{MNP}= 3 \partial _{[M} B_{NP]}$ is locally exact
and hence closed
\ben
\partial _{[Q }H_{MNP]}=0.
\een
The covariant derivative $\nabla $ defines  the usual metric-preserving
torsion-free Levi-Civita affine connection of the string metric $G_{MN}$.
It is often convenient to re-express them in terms of another
metric-preserving affine  connection whose torsion is given by the 3-form
$H_{MNP}$. This connection we call $\nabla ^+$, and acting on a vector
field $V^M$ we have
\ben
\nabla _M^+ V^N = \nabla _M V^N + { 1\over 2} H_N \, ^N \, _C V^C.
\een
The torsion $T_M \, ^N\,_ P= H_M\, ^N\,_P$.
Acting on a scalar such as the dilaton $\Phi$, $\nabla ^+$ reduces, as
does
$\nabla $, to the partial derivative
\ben
\nabla^+_M \Phi =\partial _M \Phi,
\een
However, the  ``Hessian'' is not symmetric;
in fact
\ben
2\nabla ^+_{[M} \nabla^+ _{N]} \Phi = H_M\, ^P \,_N \partial _P \Phi.
\een
Associated with $\nabla ^+$ in the usual way we have
 its curvature tensor $R^{+A} \, _{BMN}= -R^A\,_{BNM}$ and
its Ricci tensor $R^+_{BN}= R^{+A}\,_{ BAN}$, which, in the presence
of torsion, ceases to be  symmetric. 
We define the associated scalar curvature
by $R^+= G^{MN} R^+_{MN}$.  The last two equations of motion
in (\ref{betafn}) now get combined into the statement that
\ben
R^+_{MN}+ 2 \nabla ^+ _M \nabla ^+_N \Phi=0\,.
\een
The first equation becomes
\ben
4 (\nabla ^+ \Phi ) ^2 - 4 (\nabla ^+)^2 \Phi - R^+ - {1 \over 6} H^2
+ {D-26 \over 48 \pi ^2 } =0\,.
\een
In addition one must bear in mind the  torsion 3-form is closed.
This latter fact renders writing down an action principle in terms
of the string metric and the components of the connection $\Gamma^+\,
_M\,^N\, _P$ rather difficult. However one merit  of this formulation
is that it makes it easy to see  that the metric product of two solutions is
also a solution, as long as the central charges can be balanced.
It also makes it immediate that the equations of motion are solved
by a constant dilaton $\Phi$,
with the metric a product ${\Bbb E} ^{p,1}  \times G$, where $G$ is itself a
product
of  semi-simple groups carrying their Killing metrics
and where  $\nabla ^+$ is taken to be the flat parallelizing
connection obtained by right or left translation on each factor
 group. We shall encounter such solutions latter in the paper.
We shall refer to such backgrounds as WZWN ground states.

  In the string conformal frame, the action whose equations of motion imply
the vanishing of the beta functions (\ref{betafn}) is\footnote{Since,
in the special case of ten dimensions, 
this is the same action as in the NS-NS sector of superstring
theory, all the subsequent discussion of the bosonic string reductions
applies also to the NS-NS sector of superstring theory.}
\be
{\cal L} = e^{-2\Phi}\, (R\, {*\oneone} + 4 {*d\Phi}\wedge d\Phi 
-\ft12 {*H_\3}\wedge H_\3)\,.
\ee
One can also pass to the Einstein conformal frame.  Setting
\be
ds^2_{\rm string} = e^{-\ft12 a\, \phi}\, 
ds^2_{\rm Einstein}\,,\qquad
\Phi= -\fft{1}{a}\, \phi\,,
\ee
where $a^2=8/(D-2)$, gives the bosonic string Lagrangian in the
standard normalisation, in the Einstein frame.  Since we shall be
performing dimensional reductions of this theory in subsequent 
sections, we shall present it here with hats put on all the 
quantities in $D$ dimensions:
\be
\hat {\cal L} = \hat R\, {\hat * \oneone} - \ft12 {\hat * d\hat \phi}
\wedge d\hat\phi - \ft12 e^{ \hat a\, \hat\phi}\, {\hat *\hat G_\3}
\wedge \hat G_\3\,,\label{boslag}
\ee
where $\hat G_\3 = d\hat B_\2$, and the constant $\hat a$ is given by
\be
\hat a^2 = \fft{8}{D-2}\,.
\ee
The group manifold reduction, which we refer to as the ``DeWitt
reduction,'' can be viewed as a general harmonic expansion on a group
manifold $G$ of dimension $q$, in which all the lower-dimensional
fields that are associated with harmonics that are singlets under the
left action $G_L$ of the group $G$ are retained, whilst all fields
associated with non-singlet harmonics are set to zero.  This
truncation yields just a finite number of $n$-dimensional fields,
since $G_L$ acts transitively on $G$.  Crucially, the truncation is
necessarily a {\it consistent} one, meaning that setting the
non-singlet fields to zero is consistent with their own equations of
motion.  The essential point here is that despite the non-linearity of
the full system of equations, non-linear products of the retained
$G_L$ singlets can obviously never generate $G_L$ non-singlets, and
thus the retained fields cannot act as sources for the fields that are
set to zero.

\section{DeWitt Reduction of the Bosonic String}\label{dewittred}

   Let $\sigma^\a$ denote a set of left-invariant 1-forms on the 
$q$-dimensional group manifold $G$; they satisfy 
\be
d\sigma^\a = -\ft12 f^\a{}_{\beta\gamma}\, \sigma^\beta\wedge \sigma^\gamma\,,
\ee
where $f^\a{}_{\beta\gamma}$ are the structure constants of the Lie
algebra of $G$.  We shall assume throughout that $G$ is compact and
semisimple, and so the Cartan-Killing metric
\be
g_{\a\beta}\equiv -\ft12 f^\gamma{}_{\delta \alpha}\, f^\delta{}_{\gamma\beta}
\ee
is non-degenerate, and positive-definite.  It follows from the Jacobi
identity $f^\delta{}_{\lambda [\a}\, f^\lambda{}_{\beta\gamma]}=0$
that if $g_{\a\beta}$ is used to lower the
upper index on the structure constants, the resulting tensor 
$f_{\a\beta\gamma}\equiv g_{\a\delta}\, f^\delta{}_{\beta\gamma}$ is
totally antisymmetric.

\subsection{Reduction of the metric}

   We consider the standard DeWitt ansatz\footnote{It should,
perhaps, be remarked that the term ``ansatz'' is often, as here, used
inappropriately when describing dimensional reduction procedures,
since it carries the connotation that a trial substitution that might
or might not be successful is being attempted.  In fact in Kaluza
$S^1$ reductions and DeWitt group-manifold reductions there is only
one possible ``ansatz'' and there is no
possibility of failure; the ``reduction ansatz'' is nothing but an
appropriate parameterisation of the group-invariant higher-dimensional
fields in terms of lower-dimensional ones.  Since there seems to be 
no other satisfactory word that succinctly expresses the true
nature of the procedure, we shall perpetuate the use of the term
``ansatz'' despite its inappropriateness.} for reducing the 
$D=n+q$ dimensional metric $d\hat s^2$,
\be
d\hat s^2 = e^{2\a\varphi}\, ds^2 + g^{-2}\, e^{2\beta\varphi}\, h_{\a\beta}\, 
\nu^\a\,\nu^\beta\,,\label{metans}
\ee
where
\be
\nu^\a\equiv \sigma^\a - g\, A^\a\,.
\ee
Here $ds^2$ is the reduced $n$-dimensional metric, $A^\a$ are the 
Yang-Mills potentials for the gauge group $G$, and $h_{\a\beta}$ is 
a unimodular symmetric matrix parameterising the scalar degrees of
freedom.   The Yang-Mills field strengths are given by
\be
F^\a = dA^\a + \ft12 g\, f^\a{}_{\beta\gamma}\, A^\beta\wedge A^\gamma\,.
\ee
The constants $\a$ and $\beta$ are chosen to be given by
\be
\a = -\sqrt{\fft{q}{2(n-2)(n+q-2)}}
\,,\qquad \beta = -\fft{\a\, (n-2)}{q}\,.
\ee
These choices ensure that the reduction of the Einstein-Hilbert action
yields a pure Einstein-Hilbert term in $n$ dimensions, with no 
prefactor involving the breathing-mode scalar $\varphi$, and that $\varphi$ 
has a canonically-normalised kinetic term in $n$ dimensions.

    We shall choose the vielbein basis
\be
\hat e^a = e^{\a\varphi}\, e^a\,,\qquad \hat e^i = g^{-1}\, 
e^{\beta\varphi}\, \Phi^i_\a\, \nu^\a\,,
\ee
where
\be
h_{\a\beta}= \Phi^i_\a\, \Phi^i_\beta\,.
\ee
We also introduce the covariant exterior derivative $D$, whose action on
the 1-forms $\nu^\a$ is given by
\be
D\nu^\a \equiv d\nu^\a + g\, f^\a{}_{\beta\gamma}\, A^\beta\wedge \nu^\gamma
= -g\, F^\a - \ft12 f^\a{}_{\beta\gamma}\, \nu^\beta\wedge \nu^\gamma\,.
\ee

  From these expressions, we find that
\bea
d\hat e^a &=& -\a\, e^{-\a\phi}\, \del_b\phi\, \, \hat e^a\wedge \hat
e^b -\omega^a{}_b\wedge \hat e^b\,,\nn\\
d\hat e^i &=& e^{-\a\phi}\, (\Phi^{-1})^\a_j\, (D_a \Phi^i_\a)\, \hat
e^a \wedge \hat e^j + \beta \, e^{-\a\phi}\, \del_a\phi\, \hat
e^a\wedge \hat e^i - \ft12 e^{(\beta-2\a)\phi}\, F_{ab}^i\,
\hat e^a\wedge \hat e^b \nn\\
&&- \ft12 g\,e^{-\beta\phi}\, \Phi^i_\a\, (\Phi^{-1})^\beta_j\, 
(\Phi^{-1})^\gamma_k\, f^\a{}_{\beta\gamma} \,\hat e^j\wedge
 \hat e^k\,,\label{deexp}
\eea
where we have defined $F^i_{ab}\equiv \Phi^i_\a\, F^\a_{ab}$.  The 
torsion-free spin connection $\hat \omega^A{}_B$ , defined by  
$d\hat e^A = -\hat\omega^A{}_B\wedge
\hat e^B$ and $\hat\omega_{AB}=-\hat\omega_{BA}$, turns out to be
\bea
\hat \omega_{ab} &=& \omega_{ab} + \a\, e^{-\a \phi}\,
(\del_b\phi\, \eta_{ac}\, \hat e^c - \del_a\phi\, \eta_{bc}\,
\hat e^c) + \ft12 e^{(\beta-2\a)\, \phi}\, F^i_{ab}\, \hat e^i\,,\nn\\
\hat \omega_{ai} &=& -e^{-\a\phi}\, P_{a\, ij}\,  \hat e^j - \beta\,
e^{-\a\phi}\, \del_a\phi\, \hat e^i + \ft12 e^{(\beta-2\a)\, \phi}\,
  F^i_{ab}\, \hat e^b\,,\label{spincon}\\
\hat\omega_{ij} &=& e^{-\a\phi}\, Q_{a\, ij}\, \hat e^a \nn\\
&&+ \ft12g\,
e^{-\beta\phi}\, [\Phi^k_\a\, (\Phi^{-1})^\beta_i\, (\Phi^{-1})^\gamma_j
+ \Phi^j_\a\, (\Phi^{-1})^\beta_i\, (\Phi^{-1})^\gamma_k
- \Phi^i_\a\, (\Phi^{-1})^\beta_j\, (\Phi^{-1})^\gamma_k]\, 
f^\a{}_{\beta\gamma} \, \hat e^k\,,\nn
\eea
where
\be
P_{a\, ij} \equiv \ft12 [ (\Phi^{-1})^\a_i\, D_a \Phi^j_\a +
                           (\Phi^{-1})^\a_j\, D_a \Phi^i_\a]\,,
\qquad
Q_{a\, ij} \equiv \ft12 [ (\Phi^{-1})^\a_i\, D_a \Phi^j_\a -
                           (\Phi^{-1})^\a_j\, D_a \Phi^i_\a]\,.
\ee

   Defining 
\be
\hat\omega_{AB}\equiv \omega_{C\, AB}\, \hat e^C\,,\qquad
\omega_A\equiv \eta^{BC}\, \hat\omega_{B\, CA}\,,
\ee
the $(n+q)$-dimensional Einstein Hilbert Lagrangian $L=\hat e\, \hat R$ can 
be written, up to an irrelevant total derivative, as 
\be
L = \hat {\rm e}\, ( \hat \omega_{A\, BC}\,\hat\omega^{C\, AB}
  + \hat\omega^A\, \hat\omega_A)\,,\label{lagexp}
\ee
From this, one can straightforwardly determine using (\ref{spincon})
that after reduction on the group manifold $G$ the $(n+q)$-dimensional
Einstein-Hilbert action leads to an action in $n$ dimensions given in
terms of the Lagrangian $n$-form\footnote{Substitution of a Kaluza or
DeWitt reduction ansatz into a higher-dimensional Lagrangian is always
a valid procedure, provided that {\it all} the appropriate
group-invariant fields are included, and that the group is
unimodular. The resulting lower-dimensional Lagrangian yields the same
equations of motion as those that would result from substitution of
the ansatz into the higher-dimensional equations of motion.}
\bea
{\cal L}_{\rm EH}  &=& R\, {*\oneone} - \ft12 {*d\varphi}\wedge d\varphi -
{*P_{ij}}\wedge P_{ij} - \ft12 e^{2(\beta-\a)  \varphi}\,
h_{\a\beta}\, {*F^\a}\wedge F^\beta
\nn\\
&& - \ft14g^2\,  e^{2(\a-\beta)\varphi}\, (h_{\a\beta}\, h^{\gamma\delta}\, 
h^{\rho\sigma}\, f^\a{}_{\gamma\rho}\, f^\beta{}_{\delta\sigma} + 2
h^{\a\beta}\, f^\gamma{}_{\delta\a}\, f^\delta{}_{\gamma\beta})\, \,
{*\oneone}\,.\label{einstlag}
\eea

\subsection{Reduction of the 3-form field}

   We implement the group-manifold reduction at the level of the
2-form potential $\hat B_\2$, by writing
\be
\hat B_\2 = m\, g^{-3}\, \omega_\2 + B_\2 + g^{-1}\, B_{\1 \a}\wedge 
\nu^\a + \ft12 g^{-2}\, \chi_{\a\beta}\, \nu^\a\wedge \nu^\beta\,,
\label{b2red1}
\ee
where $\omega_\2$ is such that
\be
d\omega_\2 = \ft16 f_{\a\beta\gamma}\, \sigma^\a\wedge \sigma^\beta\wedge
              \sigma^\gamma\,.
\ee
We also define lower-dimensional field strengths, by writing
\be
\hat G_\3 \equiv d\hat B_\2 = G_\3 + g^{-1}\, G_{\2\, \a} \wedge \nu^\a + 
   \ft12 g^{-2}\, G_{\1\a\beta}\wedge \nu^\a\wedge \nu^\beta 
    + \ft16 g^{-3}\, 
G_{\0\a\beta\gamma}\, \nu^\a\wedge \nu^\beta\wedge \nu^\gamma\,.
\ee
It follows, therefore, that we shall have
\bea
G_\3 &=& dB_\2 + B_{\1\a}\wedge F^\a + \ft16 m\, f_{\a\beta\gamma}\, 
 A^\a\wedge A^\beta\wedge A^\gamma\,,\nn\\
G_{\2\a} &=& D B_{\1\a} + \chi_{\a\beta}\, F^\beta  + \ft12 m\, 
f_{\a\beta\gamma}\, A^\beta\wedge A^\gamma\,,\nn\\
G_{\1\a\beta} &=& D\chi_{\a\beta} + g\, f^\gamma{}_{\a\beta}\, 
B_{\1\gamma} + m\, f_{\gamma\a\beta}\, A^\gamma\,,\nn\\
G_{\0\a\beta\gamma} &=& m\, f_{\a\beta\gamma} - 3g\, \chi_{\delta[\a}\, 
  f^\delta{}_{\beta\gamma]}\,.
\eea

   It is convenient at this stage to introduce redefined
quantities as follows:
\bea
\widetilde \chi_\a &\equiv& \ft12 f_\a{}^{\beta\gamma}\, \chi_{\beta\gamma}
\,,\qquad \widetilde\chi_{\a\beta}\equiv \chi_{\a\beta}- 
f^\gamma{}_{\a\beta}\, \widetilde\chi_\gamma\,,\nn\\
\widetilde B_{\1\a} &\equiv& B_{\1\a} + m\, g^{-1}\, A_\a + g^{-1}\, 
D\widetilde\chi_\a\,,\nn\\
\widetilde B_\2 &\equiv& B_\2 - g^{-1}\, \widetilde \chi_\a\, F^\a\,,
\label{redefs}\\
\widetilde G_{\2\a} &\equiv& G_{\2\a} + m\, g^{-1}\, F_\a\,,\nn\\
\tilde\omega_\2 &\equiv&  \omega_\2 - g\, A_\a\wedge \nu^\a\,, \nn
\eea
where we have defined $F_\a\equiv g_{\a\beta}\, F^\beta$.
We also shift $\hat B_\2$ by adding to it the total derivative
$d(-g^{-2}\, \widetilde\chi_\a\, \nu^\a)$.  In terms of the new variables,
the ansatz for the potential becomes
\be
\hat B_\2 = m\, g^{-3}\, \tilde\omega_\2 + \widetilde B_\2 + 
g^{-1}\, \widetilde B_{\1\a}\wedge \nu^\a + \ft12 g^{-2}\, 
\widetilde\chi_{\a\beta}\, \nu^\a\wedge \nu\beta\,, \label{redefbs}
\ee
and the field strength $\hat G_\3 = d\hat B_\2$ is given by
\be
\hat G_\3 = G_\3 + g^{-1}\, (\widetilde G_{\2\a} -
m\, g^{-1}\, F_\a)\wedge \nu^\a + \ft12 g^{-2}\, 
G_{\1\a\beta}\wedge \nu^\a\wedge \nu^\beta 
    + \ft16 g^{-3}\, 
   G_{\0\a\beta\gamma}\, \nu^\a\wedge \nu^\beta\wedge \nu^\gamma\,,
\label{hatg3}
\ee
where
\bea
G_\3 &=& d\widetilde B_\2 + \widetilde B_{\1\a}\wedge F^\a - m\, g^{-1}\, 
      \omega_\3\,,\nn\\
\widetilde G_{\2\a} &=& D \widetilde B_{\1\a} + 
   \widetilde \chi_{\a\beta}\, F^\beta\,,\nn\\
G_{\1\a\beta} &=& D\widetilde \chi_{\a\beta} + g\, f^\gamma{}_{\a\beta}\, 
     \widetilde B_{\1\gamma}\,,\nn\\
G_{\0\a\beta\gamma} &=& m\, f_{\a\beta\gamma} - 3 g\, \widetilde 
\chi_{\delta[\a}\, f^\delta{}_{\beta\gamma]}\,,\label{redeffs}
\eea
and $\omega_\3$ is the Yang-Mills Chern-Simons 3-form, satisfying
$d\omega_\3 = F^\a\wedge F_\a$, and defined by
\be
\omega_\3 \equiv A^\a\wedge dA_\a + \ft13 g\, f_{\a\beta\gamma}\, 
A^\a\wedge A^\beta\wedge A^\gamma\,.
\ee

   It should be noted that the redefinitions for $\widetilde\chi_{\a\beta}$
and $\widetilde\chi_\a$ in (\ref{redefs}) amount to a projection of the 
scalars $\chi_{\a\beta}$, which are in the reducible antisymmetric product of 
two adjoint representations of $G$, as a sum of scalars $\widetilde\chi_\a$ 
in the adjoint of $G$, and the remaining scalars $\widetilde \chi_{\a\beta}$  
that are orthogonal to the adjoint representation.  Furthermore, it can be
seen from (\ref{redefbs}) and (\ref{redeffs}) that the scalars 
$\widetilde\chi_\a$ in the adjoint representation have disappeared entirely
from the ansatz.  In fact what has happened is that the vectors $B_{\1\a}$,
which themselves are also in the adjoint representation, have become massive
by eating the scalars $\widetilde\chi_\a$.  This becomes clear if we write
out the expression for the dimensional reduction of the $(n+q)$-dimensional 
3-form Lagrangian $-\ft12 e^{\hat a\, \hat \phi}\, 
{\hat * \hat G_\3}\wedge \hat G_\3$, 
which can easily be seen from (\ref{hatg3}) and (\ref{redeffs}) to be
given by
\bea
{\cal L}_3 &=& -\ft12 e^{\hat a\hat \phi-4\a\varphi}\, {*G_\3}\wedge G_\3 -
\ft12 e^{\hat a\hat \phi-2(\a+\beta)\varphi}\, h^{\a\beta}\, 
{*(G_{\2 \a} - \ft{m}{g}\, 
F_\a)}\wedge (G_{\2 \beta} - \ft{m}{g}\, F_\beta)\nn\\
&&
   - \ft12  e^{\hat a\hat \phi-4\beta\varphi}\, h^{\a\beta}\, 
    h^{\gamma\delta}\, {*D\widetilde\chi_{\a\gamma}}\wedge D\widetilde 
\chi_{\beta\delta} -  \ft12g^2\,   
e^{\hat a\hat \phi-4\beta\varphi}\, h^{\a\beta}\, 
    h^{\gamma\delta}\,f^\lambda{}_{\a\gamma}\, f^\sigma{}_{\beta\delta}\, 
    {*\widetilde B_{\1\lambda}}\wedge \widetilde B_{\1\sigma}\nn\\
&& -\ft1{12}  
m^2\, e^{\hat a\hat \phi + 2(\a -3\beta)\varphi}\, h^{\a_1\beta_1}\, 
  h^{\a_2\beta_2}\,  h^{\a_3\beta_3}\,
f_{\a_1\a_2\a_3}\, f_{\beta_1\beta_2\beta_3} \, {*\oneone} \label{L3red}\\
&&
-\ft14 g^2\, e^{\hat a\hat \phi + 2(\a -3\beta)\varphi}\, h^{\a_1\beta_1}\, 
  h^{\a_2\beta_2}\,  h^{\a_3\beta_3}\, f^\delta{}_{\a_2\a_3}\, 
\widetilde \chi^{\phantom {\Sigma_\Sigma}}_{\delta\a_1}\, 
(\widetilde \chi^{\phantom {\Sigma_\Sigma}}_{\lambda\beta_1}
\, f^\lambda{}_{\beta_2\beta_3} +
2 \widetilde \chi^{\phantom {\Sigma_\Sigma}}_{\lambda\beta_3}\, 
f^\lambda{}_{\beta_1\beta_2})
\, {*\oneone}\,.\nn
\eea
 We can also see from this 
Lagrangian that not only do the vectors $\widetilde B_{\1\a}$ have masses
proportional to the gauge coupling constant $g$, but so do the remaining
uneaten scalar fields $\widetilde \chi_{\a\beta}$.  

   It should be noted that if we consider the special case where the
group manifold is $G=SU(2)$, then $\widetilde\chi_{\a\beta}$ is
identically zero, since the antisymmetric product of two adjoint
representations of $SU(2)$ yields only the adjoint representation.
Thus in this special case, there are no scalars at all coming from the
reduction of $\hat B_\2$, since they are all eaten by the vectors
$B_{\1\a}$.

\subsection{The Lagrangian and gauge symmetries}

   Having obtained the DeWitt reduction of the metric and the 3-form in
the previous two subsections, we now put these results together to 
give the complete result for the DeWitt reduction of the bosonic string.

   It is convenient at this stage to perform a redefinition of the
scalars $\hat\phi$ and $\varphi$ (\ie the $(n+q)$-dimensional dilaton
and the breathing mode from the reduction), amounting to an $SO(2)$
rotation.  Specifically, we rotate these two fields $(\hat\phi,\varphi)$ to
$(\phi,\tilde\varphi)$, where $\phi$ is taken to be proportional
to the combination $\hat a\hat \phi-4\a\varphi$ that appears in the
exponential prefactor of the 3-form kinetic term in (\ref{L3red}).  In
other words, the new scalar $\phi$ can be viewed as the ``dilaton'' in 
the reduced $n$-dimensional theory.  We therefore define
\be
\phi = \fft{\hat a}{a}\, \hat\phi - \fft{4\a}{a}\, \varphi\,,\qquad
\td\varphi =  \fft{\hat a}{a}\, \varphi + \fft{4\a}{a}\,\hat\phi\,,
\ee
where
\be
a^2 = \fft{8}{n-2}\,.
\ee
The inverse relation is
\be
\hat\phi = \fft{\hat a}{a}\, \phi + \fft{4\a}{a}\, \td\varphi\,,\qquad
\varphi =  \fft{\hat a}{a}\, \td\varphi - \fft{4\a}{a}\,\phi\,,
\ee
In terms of the redefined fields, the total $n$-dimensional Lagrangian,
obtained by summing ${\cal L}_{\rm EH}$ given in (\ref{einstlag}), 
${\cal L}_3$ given in (\ref{L3red}), and the kinetic term for $\hat\phi$,
is
\bea
{\cal L}  &=& R\, {*\oneone} - 
\ft12 {*d\phi}\wedge d\phi - \ft12{*d\td\varphi}\wedge d\td\varphi-
{*P_{ij}}\wedge P_{ij} - \ft12 e^{\ft12 a\, \phi -\gamma\,\td \varphi}\,
h_{\a\beta}\, {*F^\a}\wedge F^\beta
\nn\\
&& -\ft12 e^{a\, \phi }\, {*G_\3}\wedge G_\3 -
\ft12 e^{\ft12 a\, \phi - \gamma\, \td \varphi}\, h^{\a\beta}\, 
{*(G_{\2 \a} - m\, g^{-1}\, 
F_\a)}\wedge (G_{\2 \beta} - m\, g^{-1}\, F_\beta)\nn\\
&&
   - \ft12  e^{-2\gamma\, \td\varphi}\, h^{\a\beta}\, 
    h^{\gamma\delta}\, {*D\widetilde\chi_{\a\gamma}}\wedge D\widetilde 
\chi_{\beta\delta} -  \ft12g^2\,   
e^{-2\gamma\, \td\varphi}\, h^{\a\beta}\, 
    h^{\gamma\delta}\,f^\lambda{}_{\a\gamma}\, f^\sigma{}_{\beta\delta}\, 
    {*\widetilde B_{\1\lambda}}\wedge \widetilde B_{\1\sigma}\nn\\
&&  -V\, {*\oneone}\label{fulllag}
\eea
where we have defined
\be
\gamma \equiv \sqrt{\fft{2}{q}}\,,\label{gammadef}
\ee
and the potential $V$ for the scalar fields is given by
\bea
V &=& \ft14g^2\,  
e^{-\ft12a\, \phi - \gamma\, \td\varphi}\, (h_{\a\beta}\, h^{\gamma\delta}\, 
h^{\rho\sigma}\, f^\a{}_{\gamma\rho}\, f^\beta{}_{\delta\sigma} + 2
h^{\a\beta}\, f^\gamma{}_{\delta\a}\, f^\delta{}_{\gamma\beta}) \\
&& +\ft1{12}  
m^2\, e^{-\ft12 a\, \phi -3\gamma\, \td\varphi}\, h^{\a_1\beta_1}\, 
  h^{\a_2\beta_2}\,  h^{\a_3\beta_3}\,
f_{\a_1\a_2\a_3}\, f_{\beta_1\beta_2\beta_3}  \nn\\
&&
+\ft14 g^2\, e^{-\ft12 a\, \phi -3\gamma\, \td\varphi}\, h^{\a_1\beta_1}\, 
  h^{\a_2\beta_2}\,  h^{\a_3\beta_3}\, f^\delta{}_{\a_2\a_3}\, 
\widetilde \chi^{\phantom{\Sigma_\Sigma}}_{\delta\a_1}\, 
(\widetilde \chi^{\phantom{\Sigma_\Sigma}}_{\lambda\beta_1}
\, f^\lambda{}_{\beta_2\beta_3} +
2 \widetilde \chi^{\phantom{\Sigma_\Sigma}}_{\lambda\beta_3}
\, f^\lambda{}_{\beta_1\beta_2})\,,\nn
\eea

    The Lagrangian (\ref{fulllag}) is invariant under $G$ gauge 
transformations of the Yang-Mills potentials $A_\1^\a$, with all the other 
fields transforming homogeneously, according to their Yang-Mills 
index structure.  Note, in particular, that the 1-forms $\wtd B_{\1\a}$ 
transform homogeneously in the adjoint representation of $G$.  This
should be contrasted with the original
1-forms $B_{\1\a}$ appearing in the reduction ansatz (\ref{b2red1}), which
do not transform covariantly.  Indeed, this is one reason why
it was advantageous to make the redefinition to $\wtd B_{\1\a}$ in
(\ref{redefs}).  The scalar $\phi$ deserves its appellation
``dilaton,'' since under $g_{\mu\nu}\longrightarrow \lambda^2\,
g_{\mu\nu}$, $\phi\longrightarrow \phi + 4 a^{-1}\, \log \lambda$ the
Lagrangian (\ref{fulllag}) scales uniformly as $\lambda^{n-2}$.

\section{Consistent Truncation of the Scalar Moduli}\label{truncsec}

   A careful inspection of the equations of motion that result from
the reduced Lagrangian (\ref{fulllag}) reveals that a consistent
truncation is possible in which the non-singlet scalars in the metric ansatz
(\ref{metans}) are set to zero, along with $\td\varphi$ and the scalars
$\widetilde\chi_{\a\beta}$ in the $\hat B_\2$ ansatz (\ref{redefbs}),
provided that at the same time we set the gauge bosons from the
metric reduction and the
vectors $\widetilde B_{\1\a}$ in the $\hat B_\2$ ansatz equal:
\be
h_{\a\beta}=\delta_{\a\beta}\,,\qquad \td\varphi=0\,,\qquad
\widetilde\chi_{\a\beta}=0\,,\qquad \widetilde B_{\1\a} = A_\a\,.
\label{hettrunc}
\ee
We must also set the gauge-coupling constant and the 3-form flux parameter
equal\footnote{To be precise, we can leave $g$ and $m$ unequal, and instead
set $\td\varphi=\td\varphi_0$ where $g=m\, e^{-2\gamma\, \t\varphi_0}$,
but for simplicity, and without loss of generality, we can choose the 
constant $\td\varphi_0$ to be zero and hence $g=m$.}
\be
g=m\,.
\ee
The reduction ansatz is now given simply by
\bea
d\hat s^2 &=& e^{aq/(2(n+q-2))\, \phi}\, ds_n^2 + m^{-2}\, 
     e^{-a(n-2)/(2(n+q-2))\, \phi}\, g_{\a\beta}\, \nu^\a \, \nu^\beta\,,
\nn\\
\hat B_\2 &=& m^{-2}\,  \td\omega_\2 + \widetilde B_\2 + 
              m^{-1}\, A_\a\wedge \nu^\a\,,\label{hetans}\\
\hat \phi &=& \sqrt{\fft{n-2}{n+q-2}}\, \phi\,.\nn
\eea

   The fact that this rather remarkable consistent truncation is
possible in any group manifold reduction of the bosonic string was
first discovered in \cite{het1}, where the same ansatz as
(\ref{hetans}) was presented (in the string frame rather than the
Einstein frame we are using here).  The lower-dimensional Lagrangian
that describes the truncated theory is simply obtained by imposing the
relations (\ref{hettrunc}) in (\ref{fulllag}):
\be
{\cal L} = R \, {*\oneone} - \ft12 {*d\phi}\wedge d\phi -
   \ft12 e^{a\phi}\, {*G_\3}\wedge G_\3 
-\ft12 e^{\ft12a\phi}\, {*F^\a}\wedge F^\a 
+ \ft13 q\, m^2\, e^{-\ft12 a\phi}\, 
{*\oneone}\,. \label{hetlag}
\ee

   In \cite{het1}, the higher-dimensional bosonic string theory was
augmented by the inclusion of the conformal anomaly term 
\be
{\cal L}_{\rm conf} = - k^2\, (n+q-26)\, e^{-\ft12\hat\a\, 
\hat\phi}\label{anom}
\ee
in (\ref{boslag}), and it was shown that with the same reduction
ansatz (\ref{hetans}) this additional term can cancel the scalar
potential in (\ref{hetlag}), implying that one obtains precisely the
bosonic sector of the effective action for the heterotic string. 

   In section 7 a consistent truncation of the Lagrangian
\ref{fulllag} with two singlet scalars $\Phi$ and ${\tilde \phi}$ is
given. It would be interesting to address other possible consistent
truncations with more scalars turned on and address possible domain
wall solutions there.

\section{DeWitt = Pauli  $\circ $ Kaluza }\label{dekapa}

   In section \ref{dewittred}, we described the DeWitt reduction of
the bosonic string on a group manifold $G$, in which one keeps all the
fields that are singlets under the left action of $G$.  This
reduction, as with all DeWitt reductions, is guaranteed to be
consistent, by virtue of the left-acting group invariance of the
ansatz.  By contrast, if one attempts a generalisation of the
reduction idea to a case where the internal manifold is a coset space,
such as a sphere, then aside from exceptional cases it is not possible
to perform a consistent reduction that retains a finite set of
lower-dimensional fields including all the gauge bosons of the
isometry group.  Furthermore, in those exceptional cases where such a
consistent reduction is possible, there is currently no clear
understanding, for example from group theory, as to why the
consistency is achieved.  

     In this section we shall explore some features of coset
reductions, which we refer to as ``Pauli reductions'' since the
original such example, of a 2-sphere reduction from six dimensions,
was considered by Pauli in 1953 \cite{pauli} (see \cite{strau,oraf}).
We shall describe the construction of certain classes of theory that do
admit consistent Pauli coset reductions on $G/H$. Namely,
any theory that can itself be obtained from a yet higher dimensional
``progenitor'' theory as a DeWitt group manifold reduction on $H$
admits a consistent Pauli reduction on $G/H$, for any group
$G$ that contains $H$.  The theory that results from the Pauli
reduction on $G/H$ will be the same as the theory obtained
by making a DeWitt reduction of the progenitor theory on $G$.  
The metrics induced on $G/H$ in this description are
precisely those for the consistent Pauli reduction, including the
gauge bosons of the group $G$ of isometries of $G/H$.  

   We shall demonstrate the above procedure in a variety of cases;
first for $ SU(2)/U(1)$, then $G/U(1)$ for arbitrary $G$, and finally
$G/H$ for any reductive coset.  The first two, which are examples of
the DeWitt = Pauli $\circ$ Kaluza composition, are illustrated in
Figure 1 below.  The $G/H$ case, which is a DeWitt = Pauli $\circ$
DeWitt composition, is illustrated in Figure 2.

\begin{figure}
\leavevmode\centering
\epsfxsize=0.20\textwidth
\epsfbox{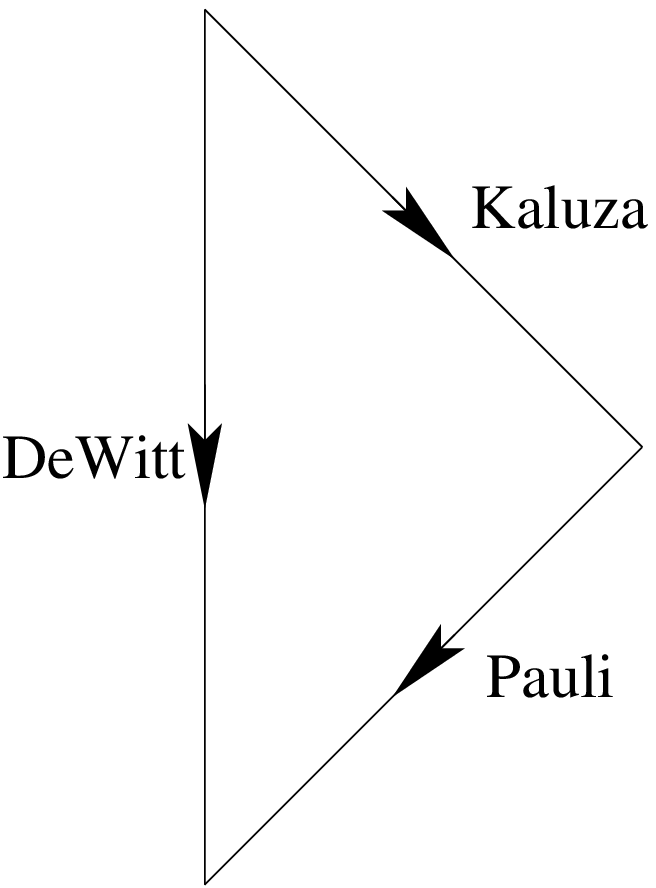}
\hspace{0.75in}
\epsfxsize=0.20\textwidth
\epsfbox{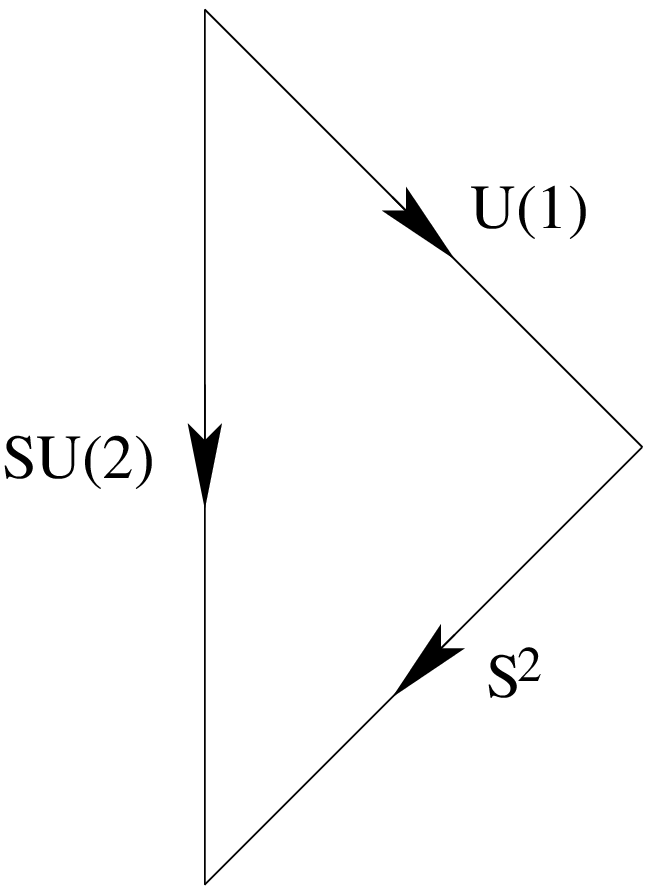}
\hspace{0.75in}
\epsfxsize=0.173\textwidth
\epsfbox{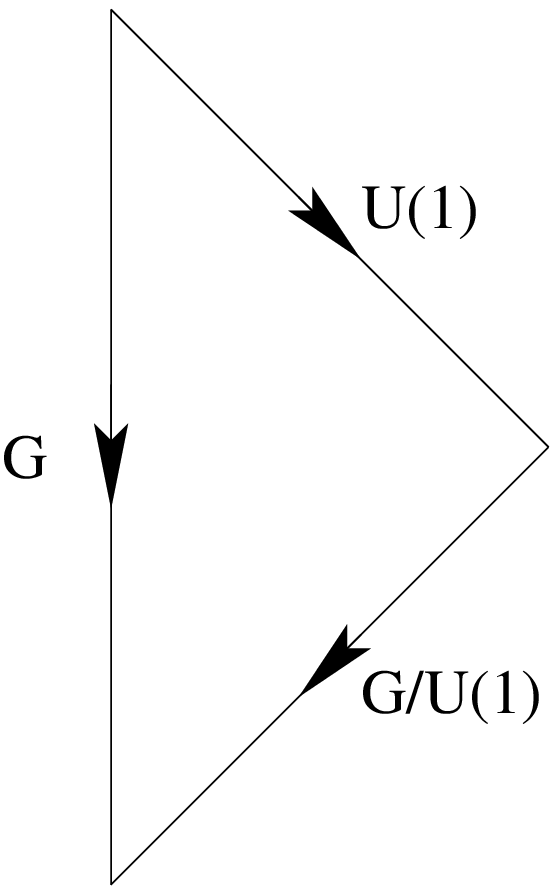}
\caption{DeWitt reduction on a group manifold $G$, composed as a Kaluza
reduction on $U(1)$ followed by a Pauli reduction on $G/U(1)$.}
\end{figure} 

\vspace{10pt}

\begin{figure}
\leavevmode\centering
\epsfxsize=0.20\textwidth
\epsfbox{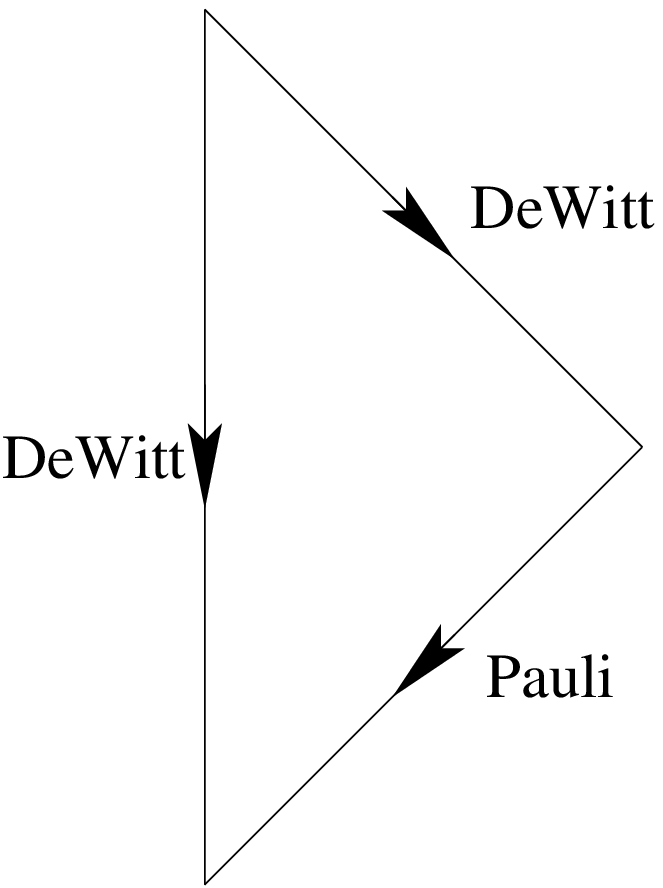}
\hspace{0.75in}
\epsfxsize=0.16\textwidth
\epsfbox{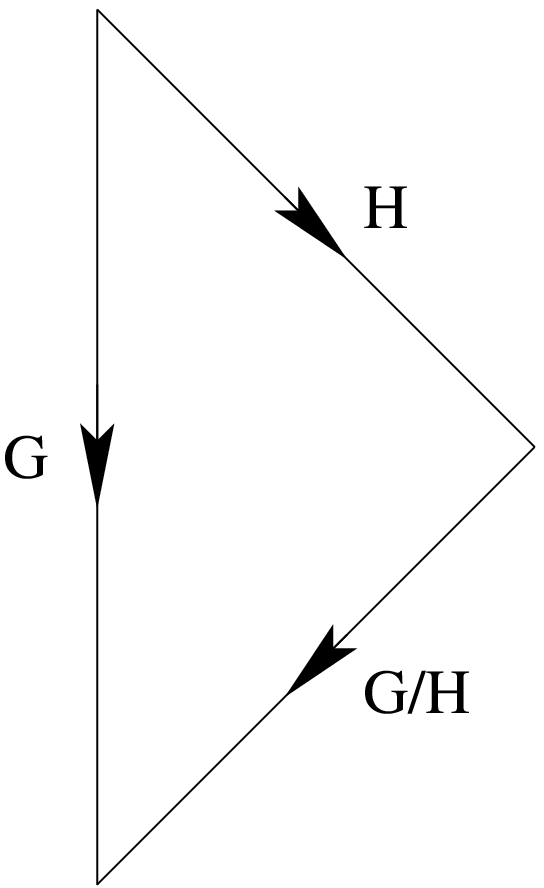}
\caption{DeWitt reduction on $G$, composed as a DeWitt reduction on
$H$ followed by a Pauli reduction on $G/H$.}
\end{figure}

\subsection{The case $SU(2)/U(1)$}\label{su2case}

   In this section, we consider the case of the DeWitt reduction on
$S^3$ of pure Einstein gravity, showing how, by instead first reducing
the Einstein theory on $S^1$, we can obtain a consistent Pauli
reduction of an Einstein-Maxwell-dilaton system on $S^2$.  This makes
contact with a result in \cite{spherered}, where it was shown that if one
starts in $D$ dimensions with the theory described by the Lagrangian
\be
{\cal L}_D = R\, {*\oneone} - \ft12 {*d\phi}\wedge d\phi
- \ft12 e^{-a\, \phi}\, {* F_\2}\wedge F_2\,,\label{einmaxdil}
\ee
where the dilaton coupling constant $a$ is given by
\be
a^2=\fft{2(D-1)}{D-2}\,,
\ee
then one can perform a consistent $S^2$ reduction that includes, in
particular, the $SU(2)$ gauge bosons associated with the isometry group
of the sphere.  This specific value of the dilaton coupling is precisely
the one that arises if $(D+1)$-dimensional pure gravity is reduced on
$S^1$ (a Kaluza reduction).  Thus the consistent $S^2$ reduction of
(\ref{einmaxdil}) derived in \cite{spherered} can be interpreted as
a consistent reduction of $(D+1)$-dimensional gravity where there
is a first step of reduction on $S^1$ followed by the reduction on $S^2$.
As we shall show, the $S^1$ fibre has a non-trivial twist, and the
whole reduction can be reinterpreted as a DeWitt reduction from $(D+1)$ 
dimensions on the
group manifold $SU(2)\sim S^3$. 

   We begin by writing down the standard DeWitt reduction of 
$(D+1)$-dimensional pure Einstein gravity on the group manifold $SU(2)$;
the metric ansatz is given by
\be
ds_{D+1}^2 = e^{2\a\varphi}\, ds_{D-2}^2 + e^{2\beta\varphi}\, 
             \wtd T_{ij}\, (\sigma^i - A^i)\, (\sigma^j - A^j)\,,
\label{d1dm2}
\ee
where $d\sigma^i = -\ft12 \ep_{ijk}\, \sigma^j\wedge \sigma^k$, and 
$\wtd T_{ij}$ denotes the unimodular matrix of scalar fields.  The
$SU(2)$ left-invariant 1-forms can be written in terms of Euler angles
$(\theta,\tau,\psi)$ as\footnote{We are using $\tau$ rather than the
more usual $\phi$ for one of the Euler angles here since $\phi$ is
already in use as a dilaton.}
\be
\sigma_1 = \cos\psi\, d\theta + \sin\psi\, \sin\theta\, d\tau\,,\quad
\sigma_2 = -\sin\psi\, d\theta + \cos\psi\, \sin\theta\, d\tau\,,\quad
\sigma_3 = d\psi + \cos\theta\, d\tau\,.
\ee
The constants $\a$ and $\beta$ in (\ref{d1dm2}) are given by
\be
\a^2 = \fft{3}{2(D-4)(D-1)}\,,\qquad \beta = -\fft{\a\, (D-4)}{3}\,.
\ee
 
    The next step is to introduce Cartesian coordinates $\mu^i$ on $\R^3$,
subject to the constraint $\mu^i\, \mu^i=1$, which defines the unit $S^2$.
We relate the $\mu^i$ to the previously-introduced Euler angles by setting
\be
\mu_1 = \sin\psi\, \sin\theta\,,\quad
\mu_2 = \cos\psi\, \sin\theta\,,\quad
\mu_3 = \cos\theta\,.
\ee
A straightforward calculation shows that we shall have
\be
\sigma^i - A^i = - \ep_{ijk}\, \mu^j\, \cD\mu^k + \mu^i\, \sigma\,,
\ee
where
\be
\sigma\equiv d\tau + \cos\theta\, d\psi - \mu^i\, A^i\,,\qquad
\cD\mu^i \equiv d\mu^i + \ep_{ijk}\, A^j\, \mu^k\,.
\ee
After some involved but mechanical manipulations, we find that we can
write (\ref{d1dm2}) as 
\be
d\hat s_{D+1}^2 = e^{2\a\varphi}\, ds_{D-2}^2 + e^{2\beta\varphi}\, 
\wtd\Delta^{-1}\, 
\wtd T^{-1}_{ij}\, 
\cD\mu^i\, \cD\mu^j + e^{2\beta\varphi}\, \Delta\, (d\tau + A)^2\,,
\label{d1dm22}
\ee
where
\bea
A &=& \cos\theta\, d\psi - \mu^i\, A^i - \wtd\Delta^{-1}\, \wtd T_{ij}\, 
\ep_{ik\ell}\, \mu^j\, \mu^k\, \cD\mu^\ell\,,\nn\\
\wtd\Delta &\equiv& \wtd T_{ij}\, \mu^i\, \mu^j\,.
\eea

    Since $\tau$ is an isometry direction, we can now perform a
standard Kaluza reduction of (\ref{d1dm22}) on the circle
parameterised by $\tau$, by writing it as
\be
d\hat s_{D+1}^2 = e^{2\td\a\phi}\, ds_{D}^2 + g^{-2}\, e^{2\td\beta\phi}\, 
              (d\tau + A)^2\,,\label{kaluzared}
\ee
where
\be
\td \a^2 =\fft1{2(D-1)(D-2)}\,,\qquad \td\beta = - (D-2)\, \td\a\,.
\ee
Since we are starting from pure Einstein gravity in $(D+1)$ dimensions, 
it follows that the metric $ds_D^2$, the vector potential $A$ and the
dilaton $\phi$ will satisfy the equations of motion for the $D$-dimensional
Einstein-Maxwell-dilaton system described by the Lagrangian (\ref{einmaxdil}).

   Comparing with (\ref{d1dm22}), we see that this gives the following 
$S^2$ reduction ansatz from $D$ dimensions to $(D-2)$ dimensions:
\bea
ds_D^2 &=& e^{2\a\varphi - 2\td\a \phi}\, ds_{D-2}^2 + g^{-2}\, 
e^{2\beta\varphi- 2\td\a\phi}\, \wtd\Delta^{-1}\, \wtd T^{-1}_{ij}\, 
D\mu^i\, D\mu^j\,,\nn\\
A &=&  \cos\theta\, d\psi - \mu^i\, A^i - \wtd\Delta^{-1}\, \wtd T_{ij}\, 
\ep_{ik\ell}\, \mu^j\, \mu^k\,D\mu^\ell\,,\label{s2pauli3}\\
e^{2\td\beta\phi} &=& e^{2\beta\varphi}\, \wtd\Delta\,.\nn
\eea
It is straightforward to re-express this reduction ansatz as
\bea
ds_D^2 &=& Y^{\ft1{D-2}}\, \Big( \Delta^{\ft1{D-2}}\, ds_{D-2}^2 + g^{-2}\, 
\Delta^{-\ft{D-3}{D-2}}\, T_{ij}^{-1}\, D\mu^i\, D\mu^j\Big)\,,\nn\\
e^{\sqrt{\ft{2(D-2)}{D-1}}\, \phi} &=& \Delta^{-1}\, Y^{\ft{D-3}{D-1}}
\,,\label{s2pauli}\\
A &=&  \cos\theta\, d\psi - \mu^i\, A^i - \Delta^{-1}\, T_{ij}\, 
\ep_{ik\ell}\, \mu^j\, \mu^k\,D\mu^\ell\,,\nn
\eea
where we have absorbed the breathing-mode scalar $\varphi$ introduced
in (\ref{d1dm2}) into the matrix $T_{ij}$, defined in terms of the
unimodular matrix $\wtd T_{ij}$ by $T_{ij}\equiv Y^{1/3}\, \wtd
T_{ij}$, and $Y$ is given by $Y=e^{(D-1)\, \a\, \varphi}$.  After a
somewhat lengthy calculation, we find that the field strength $F=dA$
is then expressible as
\bea
F &=& \ft12\ep_{ijk}\, \Big( g^{-1}\, U\, \Delta^{-2}\, \mu^i\, D\mu^j \wedge
D\mu^k - 2 g^{-1}\, \Delta^{-2}\, D\mu^i \wedge D T_{j\ell}\, T_{km}\,
\mu^\ell\, \mu^m\Big) \nn\\
&&- \Delta^{-1}\, T_{ij}\, \mu^i\, F^j\,.
\eea
The $S^2$ reduction that we have derived here is precisely the 
consistent Pauli reduction ansatz derived first in \cite{spherered}.

     It should be emphasised that although we have focused here on the 
$S^3$ DeWitt reduction of pure Einstein gravity in $(D+1)$ dimensions,
and its consequent reinterpretation as a Pauli reduction of the $D$-dimensional
Einstein-Maxwell-dilaton system, we could just as well begin with {\it any}
theory in $(D+1)$ dimensions.  That theory, reduced on a circle, will then,
by the same arguments, yield a $D$-dimensional theory that can necessarily
be consistently Pauli-reduced on $S^2$.  Likewise, the results in the 
next subsection extend to show that this $D$-dimensional theory can be
consistently Pauli-reduced on {\it any} coset $G/U(1)$.  

   An example of a $D$-dimensional theory that can be obtained from a
circle reduction is type IIA supergravity, since it comes from the
$S^1$ reduction of eleven-dimensional supergravity.  Thus, for
example, we are guaranteed to be able to find a consistent Pauli
reduction of type IIA supergravity on $S^2$.  In fact the resulting
eight-dimensional theory will be precisely the same $SU(2)$-gauged
supergravity as the one obtained in \cite{salsez8} by performing the
DeWitt reduction of eleven-dimensional supergravity on $S^3=SU(2)$.

\subsection{The case $G/U(1)$}\label{gcase}

    In this section, we show how the previous discussion can be 
generalised to the case of any group $G$ factored by $U(1)$.  
We begin by introducing generators $T_a$ for the 
Lie algebra of $G$, and defining the left-invariant and right-invariant 
1-forms of $G$ by
\be
\lambda^a\, T_a = V^{-1}\, dV\,,\qquad \rho^a\, T_a = dV\, V^{-1}\,,
\label{lrdef}
\ee
where $V\in G$.
It is evident that the matrix $U^a{}_b$ defined by
\be
T_a\, U^a{}_b = V^{-1}\, T_b\, V\label{udef}
\ee
has the properties
\be
U^a{}_c\, U_b{}^c = U^c{}_b\, U_c{}^a = \delta^a_b\,,
\ee
where indices are raised and lowered using the Cartan-Killing metric
$g_{ab}= - \tr(T_a\, T_b)$.  We can use $U^a{}_b$ to relate the left-invariant
and right-invariant 1-forms:
\be
\lambda^a = U^a{}_b\, \rho^b\,,\qquad \rho^a = U_b{}^a\, \lambda^b\,.
\ee

    We now split the generators as $T_a=(T_0, T_i)$, where 
$T_0$ is the generator of the relevant $U(1)$ subgroup, and
parameterise group elements $V\in G$ in the form
\be
V = e^{\tau\, T_0}\, \wtd V\,,
\ee
where $\wtd V$ parameterises elements in the coset $G/U(1)$.  Thus,
in particular, $\wtd V$ is independent of the coordinate $\tau$ on the 
$U(1)$ circle.  From (\ref{lrdef}), we find
\bea
\rho^a\, T_a &=& d\tau\, T_0 + e^{\tau\, T_0}\, d\wtd V\, \wtd V^{-1}\, 
e^{-\tau\, T_0}\,,\nn\\
  &=& (d\tau + \omega)\, T_0 + \rho^i\, T_i\,,\label{rhoexp}
\eea
where $d\wtd V\, \wtd V^{-1} = \omega\, T_0 + \td\rho^i\, T_i$ and
$\rho^i\, T_i = e^{\tau\, T_0}\, T_i\, e^{-\tau\, T_0}\, \td\rho^i$.
Note that $\omega$ and $\td\rho^i$ are independent of $\tau$.  

   From (\ref{lrdef}), the left-invariant 1-forms $\lambda^a$ are given 
by
\bea
\lambda^a\, T_a &=& d\tau\, \wtd V^{-1}\, T_0\, \wtd V + \wtd V^{-1}\, d
\wtd V\,,\nn\\
&=& (U^a{}_0\, \rho^0 + U^a{}_i\, \rho^i)\, T_a\,,\label{lamexp}
\eea
where the second line follows from (\ref{udef}).  Now, it is evident
by setting $b=0$ in (\ref{udef}) that we have
\be
U^a{}_0 = V^{-1}\, T_0\, V = \wtd V^{-1}\, T_0\, \wtd V\,,
\ee
and hence that $U^a{}_0$ is independent of $\tau$.  It is also evident 
that
\be
U^a{}_i\, \rho^i\, T_a =  V^{-1}\, (\rho^i\, T_i)\, V
=\wtd V^{-1}\, e^{-\tau\, T_0}\, (\rho^i\, T_i)\, e^{\tau\, T_0}\, \wtd V
 = \wtd V^{-1}\, (\td\rho^i\, T_i)\, \wtd V\,,
\ee
which, from the already-established $\tau$-independence of $\td\rho^i$
implies that $U^a{}_i\, \rho^i$ is independent of $\tau$.   The upshot
of these observations is that, from (\ref{rhoexp}) and (\ref{lamexp}), 
we have
\be
\lambda^a = U^a{}_0 \, (d\tau+\omega) + U^a{}_i\, \rho^i\,,\label{lamfin}
\ee
where $U^a{}_0$, $\omega$ and $U^a{}_i\, \rho^i$ are all independent of
$\tau$, and of course $\rho^i$ is orthogonal to $\del/\del\tau$.  

   Armed with these preliminaries, we can now consider the standard
DeWitt metric reduction on the group manifold $G$:
\be
d\hat s^2 = e^{2\a\, \varphi}\, ds^2 + g^{-2}\, e^{2\beta\, \varphi}\, 
T_{ab}\, (\lambda^a - A^a)\, (\lambda^b- A^b)\,,\label{dewitt7}
\ee
(It is understood here that $T_{ab}$ is taken to be unimodular; 
also, without loss of
generality we are setting the gauge coupling constant $g=1$.)
Substituting (\ref{lamfin}) into this, we obtain
\be
d\hat s^2 = e^{2\a\, \varphi}\, ds^2 + 
e^{2\beta\, \varphi}\, \Delta\, (d\tau + \omega - U_a{}^0\, A^a)^2 +
 2 T_{ab}\, U^a{}_0\, (d\tau+\omega - U_a{}^0\, A^a)\, h^b+ 
T_{ab}\, h^a\, h^b \,,\label{dwans3}
\ee
where we have defined
\be
h^a \equiv U^a{}_i\, (\rho^i - U_b{}^i\, A^b)\,,\qquad \Delta\equiv 
T_{ab}\, U^a{}_0\, U^b{}_0\,.
\ee
Note that since $U^a{}_i\, U_b{}^i= U^a{}_c\, U_b{}^c - U^a{}_0\, U_b{}^0
=\delta^a_b - U^a{}_0\, U_b{}^0$, and we have already established that
$U^a{}_0$ is independent of $\tau$, it follows that $U^a{}_i\,
U_b{}^i$ is also independent of $\tau$, and hence so is $h^a$.  Thus by
taking all the fields $ds^2$, $T_{ab}$, $\varphi$ and $A^a$ to be
independent of $\tau$, we have a $U(1)$ isometry generated by
$\del/\del\tau$, which can be used for a standard Kaluza reduction.
To do this, we complete the square in (\ref{dwans3}), giving
\bea
d\hat s^2 &=&  e^{2\a\,\varphi}\, ds^2 + e^{2\beta\, \varphi}\,
\Big(d\tau + \omega - U_a{}^0\, A^a + \Delta^{-1}\, 
T_{ab}\, U^a{}_0\, h^b \Big)^2 \nn\\
&& + e^{2\beta\, \varphi}\, 
(T_{ab} - \Delta^{-1}\, T_{ac}\, 
T_{bd}\, U^c{}_0\, U^d{}_0)\, h^a\, h^b \,.\label{u1isom}
\eea

   If we start from pure Einstein gravity in $(D+1)$ dimensions, 
and reduce on the $U(1)$ isometry generated by $\del/\del\tau$ using the
standard reduction ansatz (\ref{kaluzared}), 
we therefore find that the DeWitt reduction of the $(D+1)$-dimensional 
theory on $G$ can be reinterpreted as a Pauli reduction on the
coset $G/U(1)$ of the $D$-dimensional Einstein-Maxwell-dilaton
theory (\ref{einmaxdil}), with the reduction ansatz given by
\bea
ds_D^2 &=& (e^{2\beta\, \varphi}\, \Delta)^{-\ft1{D-2}}\, 
(  e^{2\a\,\varphi}\, ds^2 +  e^{2\beta\, \varphi}\, 
(T_{ab} - \Delta^{-1}\, T_{ac}\, 
T_{bd}\, U^c{}_0\, U^d{}_0\, h^a\, h^b)\,,\nn\\
A_\1 &=& \omega -  U_a{}^0\, A^a + \Delta^{-1}\, T_{ab}\, U^a{}_0\, h^b 
\,,\label{paulig}\\
e^{2\td\beta\, \phi} &=& e^{2\beta\, \varphi}\, \Delta\,.\nn
\eea
The automatic consistency of the DeWitt reduction on $G$ ensures the 
consistency of this Pauli reduction on $G/U(1)$.

   It is instructive to make contact with our results for the case
$SU(2)/U(1)$ in section \ref{su2case}.  To do this, we 
note that the terms in the metric reduction (\ref{paulig}) in the
coset directions can be written as
\be
(e^{2\beta\, \varphi}\, \Delta)^{-\ft1{D-2}}\,  e^{2\beta\, \varphi}\, 
\Delta^{-1}\, (T_{ab}\, T_{cd}  - T_{ac}\, 
T_{bd}) \, U^c{}_0\, U^d{}_0\, h^a\, h^b\,.
\ee
Since $T_{ab}$ is $3\times 3$ and unimodular, we have 
\be
T_{ab}\, T_{cd} - T_{ac}\, T_{bd} = \ep_{ace}\, \ep_{bdf}\, T^{-1}_{ef}\,.
\ee
Defining $\td h^a= \ep_{abc}\, h^b\, U^c{}_0$, we find that the 
metric reduction in (\ref{paulig}) can be written in this special 
case as
\be
ds_D^2 =  (e^{2\beta\, \varphi}\, \Delta)^{-\ft1{D-2}}\, 
(e^{2\a\,\varphi}\, ds^2 +  e^{2\beta\, \varphi}\, \Delta^{-1}\, 
T_{ab}^{-1}\, \td h^a\, \td h^b)\,.
\ee
This can be seen to be equivalent to the metric reduction in
(\ref{s2pauli3}), with $U^a{}_0=\mu^a$ and $\td h^a=\cD\, \mu^a$.  
Likewise, the reductions for $A_\1$ and $\phi$ in (\ref{paulig})
coincide in this case with the expressions in (\ref{s2pauli3}).

\subsection{The case $G/H$}\label{ghcase}

   It should now be clear that the procedure we have described
in section \ref{gcase} for $G/U(1)$ can be generalised 
to any coset $G/H$.  This can easily seen if we parameterise
the group element $V\in G$ as 
\be
V= h\, k\,,
\ee
where $h\in H$ and $k\in {G/H}$.  The DeWitt reduction on $G$
is written in terms of $T_{ab}\, (\lambda^a-A^a)\, (\lambda^b-A^b)$,
and so one needs to evaluate
\be
T_a\, (\lambda^a - A^a) = V^{-1}\, dV - A\,,
\ee
where $A\equiv T_a\, A^a$, and $T_a$ are the generators of $G$.  One then
has
\bea
T_a\, (\lambda^a - A^a) &=& k^{-1}\, h^{-1}\, dh\, k + k^{-1}\, dk - A\nn\\
&=& k^{-1}\, (h^{-1}\, dh + dk\, k^{-1} - k\, A\, k^{-1} )\, k
\label{dhdk}
\,.
\eea
  The terms enclosed within the parentheses in the second line
of (\ref{dhdk}) can be split into the contribution $h^{-1}\, dh +
(dk\, k^{-1})_\| + (k\, A\, k^{-1})_\|$ parallel to the $h$ fibres,
and the contribution $(dk\, k^{-1})_\perp + (k\, A\, k^{-1})_\perp$
perpendicular to the fibres (\ie in the $G/H$ base).  

    We split the generators
$T_a$ of $G$ as $T_a=(T_\a\,, T_i)$, where $T_\a$ generate the subgroup
$H$.  One can parameterise elements in $H$ as
\be
h= e^{\tau_\a\, T_\a}\,.
\ee
The left-invariant 1-forms $\lambda^a$ for $G$ are then given by
\be
\lambda^a\, T_a = \Lambda^\a\, k^{-1}\, T_\a\, k + k^{-1}\, dk\,,
\label{lamh}
\ee
where $\Lambda^\a$ are the left-invariant 1-forms on the subgroup
$H$:
\be
\Lambda^\a\, T_\a = h^{-1}\, dh\,.
\ee
It is clear from (\ref{lamh}) that one can substitute the $\lambda^a$
into (\ref{dewitt7}), and then perform a DeWitt reduction on the
fibres of the group manifold $H$, by completing the square on the
terms involving $\Lambda^\a$.  Thus one has a ``DeWitt = Pauli $\circ$ 
DeWitt'' interpretation, for any group $G$ with subgroup $H$.

    If the coset is reductive, meaning in particular that $[H,K] = K$,
we can obtain rather elegant explicit formulae for the reduction, which
give a natural generalisation of the expressions in section
\ref{gcase}.\footnote{In practice the cosets occurring in 
dimensional reductions on compact spaces are always reductive.}   
We derive these formulae in Appendix A.

\section{Pauli Reductions of the Bosonic String}

\subsection{Introduction}

     In section \ref{dewittred}, we discussed the standard DeWitt
procedure for performing a group-manifold dimensional reduction,
applied to the specific case of the $D$-dimensional bosonic string.
As always in the DeWitt reduction, the consistency of the procedure is
guaranteed by virtue of the fact that the ansatz is invariant under
the transitively acting left action of the group $G$ on the reduction
manifold.  In section \ref{truncsec}, we discussed the details of the
consistent truncation of this group manifold reduction that can be
performed in the special case of the bosonic string.  Namely, one can
consistently set to zero all the scalar modulus fields, provided that 
at the same time one equates the vectors coming from the reduction of the
2-form potential and the vectors coming from the reduction of the metric.
This consistent truncation, which was first discovered in \cite{het1},
results in a lower-dimensional theory comprising just gravity, the
gauge bosons of $G$, and the dilaton.  The fact that this truncation 
can be performed consistently depends on specific features of the bosonic
string, and there is no obvious group-theoretic explanation for it. The
calculations are sufficiently straightforward in this case, however, that
one may regard the explicit demonstration of the consistency as an 
adequate, if unilluminating, explanation in its own right.

    A more subtle situation arises if we consider more general
possibilities for dimensional reduction of the bosonic string on a
group manifold.  Since the bi-invariant metric on the manifold $G$ has
isometry group $G_L\times G_R$, one can enquire whether it is possible to
perform a consistent reduction in which the gauge bosons of the entire
$G_L\times G_R$ isometry group are retained.\footnote{As usual, we are
addressing ourselves to situations where only a finite total number of
lower-dimensional fields are to be retained in the reduction ansatz.}
If one were considering such a possibility in the reduction of a
generic higher-dimensional theory, the answer would certainly be ``no.''
However, the special features of the bosonic string that have already
been seen to play a role in section \ref{truncsec} suggest that further
remarkable consistent truncations may be possible.  Indeed, persuasive
evidence was obtained in \cite{het2} which implies that a consistent
reduction of the bosonic string that retains the gauge bosons of $G\times G$ 
should be possible.  Specifically, by considering a reduction ansatz that
was exact in the gauge-boson sector, and correct up to linear order in scalar
fields, it was shown that a highly non-trivial potential obstacle to the
consistency of the reduction was avoided, as a consequence of certain
conspiracies between contributions from the metric and the 2-form 
reduction ans\"atze.  As a result, it was conjectured in \cite{het2}
that there always exists a consistent reduction of the bosonic string
on a group manifold $G$, in which the lower-dimensional fields comprise
the metric, the dilaton, the gauge bosons of $G_L\times G_R$, and 
scalar fields in the representation (Adjoint($G_L$), Adjoint($G_R$)).
In the terminology of the present paper we refer to this 
as a Pauli reduction, since in the spirit of Pauli's proposed $S^2$ reduction
scheme it is a case where gauge bosons for the entire isometry group of
the reduction manifold are obtained.

    Further supporting evidence for the consistency of this Pauli
reduction comes from considerations discussed in \cite{spherered}.  In
that paper, a general argument that yields a {\it necessary} condition
for the existence of a consistent reduction was introduced.  The
argument is as follows.  Suppose a $D$-dimensional theory is such that
when dimensionally reduced on the $q$-torus it yields a theory in
$(D-q)$ dimensions with a global symmetry group $P$, whose maximal
compact subgroup is $Q$.  If instead the $D$-dimensional theory is
reduced on a compact manifold $M_q$, then a necessary condition for
there to exist a consistent such reduction is that the gauge bosons in
the reduction ansatz, coming from the isometry group of $M_q$, must be
contained within the maximal compact subgroup $Q$ of the toroidal
reduction.  The argument for this is that one can always take a limit
where the
scale-size of $M_q$ is sent to infinity in the putative consistent reduction
on $M_q$, and in this limit the curved manifold $M_q$ effectively
approaches the flat torus $T^q$.  Conversely, the lower-dimensional
theory that one would obtain from the $M_q$ reduction can be viewed as
a gauging of the theory coming from the $T^q$ reduction.  However, the
process of gauging requires that one gauge a subgroup of the maximal
compact subgroup $Q$ of the global symmetry of the ungauged
toroidally-reduced theory.  Thus it follows that one cannot obtain a
theory from the $M_q$ reduction whose gauge group lies outside the
maximal compact subgroup $Q$.

   If we now apply this argument to the case of the bosonic string, we
know that after a Kaluza-type dimensional reduction on $T^q$ we obtain
a theory with scalars in the coset $O(q,q)/(O(q)\times O(q))$, and thus
the maximal compact subgroup $Q$ is $O(q)\times O(q)$.  If we now
instead consider reducing the bosonic string on a group manifold $G$
whose dimension is dim($G$) $=q$, then we see that the necessary 
condition for the consistency of a Pauli reduction retaining the gauge
bosons of $G\times G$ is that we should have
\be
G\times G \subset   O(q)\times O(q)\,.
\ee
This is in fact the case for any group $G$, since, as is well known,
there is a natural embedding of any compact group $G$ in the 
orthogonal group $O(q)$ where $q$ is the dimension of $G$.  Thus the 
necessary condition for the existence of a consistent Pauli reduction
of the bosonic string is satisfied for any group manifold $G$.

   One can easily see that special properties of the bosonic string
are playing a crucial role here.  If we were instead to entertain the
possibility of a consistent Pauli reduction of a generic theory such
as $D$-dimensional pure gravity, the analogous test would be to see
whether $G\times G$ is contained in the denominator group of the
$P/Q=SL(q,\R)/O(q)$ or $P/Q=GL(q,\R)/O(q)$ scalar coset that would
arise in such cases.  Clearly the answer is ``no,'' and so one could
not hope to retain more than the single copy of $G$ that arises in a
DeWitt reduction.

\subsection{Pauli reductions of the bosonic string on $S^3$ and group 
manifolds}

    In fact an example of a consistent Pauli reduction of the bosonic 
string on a group manifold $G$ is already known in the literature, for the
case of $G=SU(2)\sim S^3$ \cite{spherered}.  A complete and explicit 
reduction ansatz for this example was obtained in \cite{spherered}, which
yields a reduced theory that includes all six gauge bosons of the
$SO(4)$ isometry group of the round 3-sphere, together with ten scalar
fields.  The reduction ansatz was given in the Einstein frame in 
\cite{spherered}, but for our present purposes it is simpler to work
in the string frame.   After making the appropriate redefinitions,
the ansatz becomes
\bea
d\hat s_D^2 &=& ds_{D-3}^2 + g^{-2}\, \Delta^{-1}\, Y\, ^{1/2}\, 
T^{-1}_{ij}\, 
\cD\mu^i\, \cD\mu^j\,,\nn\\
e^{-2 \hat\phi/\hat a} &=& \Delta^{-1}\, Y^{(D-4)/4}\,,
\label{phians}\nn\\
\hat F_\3 &=& F_\3 + \ft16\, \ep_{i_1 i_2 i_3 i_4}\,
\Big( g^{-2}\, U\, \Delta^{-2}\,
  \cD\mu^{i_1}\wedge \cD\mu^{i_2} \wedge \cD\mu^{i_3}\,
\mu^{i_4} \label{fans}\\
&&\!\!\!
 - 3g^{-2}\, \Delta^{-2} \,
\cD\mu^{i_1} \wedge \cD\mu^{i_2}\wedge \cD T_{i_3 j}\,
T_{i_4 k}\, \mu^j\, \mu^k  - 3g^{-1}\, \Delta^{-1}\, F_\2^{i_1 i_2} \wedge
\cD\mu^{i_3}\, T_{i_4 j}\, \mu^j \Big)\,,\nn
\eea
where
\be
\mu^i \, \mu^i = 1\,,\qquad \Delta = T_{ij}\, \mu^i\, \mu^j\,,\qquad
U = 2 T_{ik}\, T_{jk}\, \mu^i\, \mu^j - \Delta \, T_{ii}\,,\nn\\
\ee
the indices $i,j,\ldots$ range over 4 values, which we take to be
$(0,1,2,3)$, and we define $Y\equiv \det(T_{ij})$.  
The gauge-covariant exterior derivative $\cD$ is defined so that
\be
\cD\mu^i = d\mu^i + g\, A_\1^{ij}\, \mu^j\,,\qquad
\cD T_{ij} = dT_{ij} + g\, A_\1^{ik}\, T_{kj} + g\, A_\1^{jk}\, T_{ik}\,,
\label{gaugecov}
\ee
where $A_\1^{ij}$ denotes the $SO(4)$ gauge potentials coming from the
isometry group of the 3-sphere, and
\be
F_\2^{ij} = dA_\1^{ij} + g\, A_\1^{ik}\wedge A_\1^{kj}\,.
\label{fieldstrength}
\ee
Note that $Y$ parameterises the lower-dimensional dilaton $\phi$,
namely $Y=e^{-a\, \phi}$ with $a^2 = 8/(D-5)$.  The lower-dimensional
metric appearing in (\ref{phians}) is written in its string frame;
this is related to the Einstein frame by 
\be
ds_{D-3}^2({\rm string}) =
Y^{1/2}\, ds_{D-3}^2({\rm Einstein})= e^{-\ft12 a\, \phi}\,
ds_{D-3}^2({\rm Einstein})\,.
\ee

   We would like to generalise the above $S^3$ Pauli reduction results to
obtain an ansatz for the Pauli reduction of the bosonic string on an
arbitrary group manifold.  For the reduction of the field strength
$\hat F_\3$, this is a complicated problem that we have not yet
succeeded in solving.  For the metric reduction ansatz, on the other
hand, there does seem to be a natural conjecture for its form.  We can
see this by first noting that in the case of $S^3$, the metric
reduction ansatz in (\ref{phians}) can be written as
\be
d\hat s_D^2 = ds_{D-3}^2 + g^{-2}\, \wtd\Delta^{-1}\, \wtd T_{ij}^{-1}\, 
\cD\mu^i\, \cD\mu^j\,,\label{gggg}
\ee
where $\wtd T_{ij}\equiv Y^{-1/4}\, T_{ij}$ is unimodular, and
$\wtd\Delta \equiv \wtd T_{ij}\, \mu^i\, \mu^j$.  After some algebra, 
we find that the metric (\ref{gggg}) has the inverse
$(\del/\del \hat s)^2\equiv \hat g^{MN}\, \del_M\otimes \del_N$, with
\be
\Big(\fft{\del}{\del \hat s}\Big)^2 = \ft12 g^2\, \wtd T_{ik}\, 
\wtd T_{j\ell}\, K^{ij} 
\otimes K^{k\ell }  
+ g^{\mu\nu}\, (\del_\mu + \ft12 g\, A_\mu^{ij}\, K^{ij})\otimes
 (\del_\nu + \ft12 g\, A_\nu^{k\ell}\, K^{k\ell})\,,\label{inversemet}
\ee
where $g^{\mu\nu}$ is the inverse of the lower-dimensional spacetime
metric, and $K^{ij}$ denotes the $SO(4)$ Killing vectors,
\be
K^{ij} = \mu^i\, \fft{\del}{\del \mu_j} - \mu^j\, \fft{\del}{\del\mu^i}\,.
\label{snkv}
\ee
In fact (\ref{inversemet}), which can be proved relatively easily by
diagonalising $\wtd T_{ij}$ at a point, admits an immediate
generalisation to any sphere $S^n$ for arbitrary $n$, simply by
allowing the index range on $\mu^i$ to be $0\le i\le n$.

    It should be emphasised that in writing the inverse metric as in
(\ref{inversemet}), we are using an ``overcomplete'' basis of vector fields,
since on $S^n$ we have $\ft12 n(n-1)$ Killing vectors $K^{ij}$ on a 
space that is only $n$-dimensional.  The advantage of using this 
overcomplete basis, however, is that the expression (\ref{inversemet})
does not require the use of the $S^n$-dependent quantity $\wtd\Delta=
\wtd T_{ij}\, \mu^i\, \mu^j$ that appears in the metric ansatz in 
(\ref{phians}).  

    The next step is to write the inverse metric (\ref{inversemet}) for 
the $S^3$ reduction in terms of self-dual and anti-self-dual 
combinations of the $SO(4)$ Killing vectors, since these are the 
combinations that generate the right and left actions of $SU(2)$.
This will cast the $S^3$ inverse-metric ansatz into a form that admits
a natural generalisation to the case of any group manifold $G$.
The right and left combinations for $S^3$, denoted by $R^a$ and $L^{a}$, 
are given by
\be
R^a = \ft12 \eta^a_{ij}\, K^{ij}\,,\qquad 
L^{a} = \ft12 \bar\eta^{a}_{ij}\, K^{ij}\,,
\ee
where $\eta^a_{ij}$ and $\bar\eta^{a}_{ij}$ are the self-dual
and anti-self-dual 't Hooft tensors, whose components are defined by
\be
\eta^a_{0b} = -\delta^a_b\,,\quad \eta^a_{bc} = -\ep_{abc}\,,\quad
\bar\eta^a_{0b} = -\delta^a_b\,,\quad \eta^a_{bc} = \ep_{abc}\,.
\ee
Defining associated sets of right and left $SU(2)$ Yang-Mills fields
by 
\be
A_\mu^a = \ft14 \eta_{ij}^a\, A_\mu^{ij}\,,\qquad B_\mu^a
=\ft14 \bar\eta_{ij}^a\, A_\mu^{ij}\,,
\ee
we can rewrite (\ref{inversemet}) as
\bea
\Big(\fft{\del}{\del \hat s}\Big)^2 &=& \ft12 g^2\, 
M_{ab}\ R^a\otimes R^b + 
\ft12 g^2\, N_{ab}\, L^{a}\otimes L^{b} + 
\ft12 g^2\, P_{a b}\, 
  (R^a\otimes L^{b}+ L^{b}\otimes R^a)\nn\\
&&+ g^{\mu\nu}\, (\del_\mu +  g\, A_\mu^a\, R^a + g\, 
B_\mu^{a}\, L^{a})
\otimes
 (\del_\nu + g\, A_\nu^b\, R^b + g\, B_\nu^{b}\, L^{b}
)\,,\label{inversemet2}
\eea
where $M_{ab}$, $N_{ab}$ and $P_{ab}$ are given in terms of 
$\wtd T_{ij}$ by
\be
M_{ab} = \ft14 \wtd T_{il}\, \wtd T_{j\ell}\, \eta^a_{ij}\, 
\eta^b_{k\ell}\,,\quad
N_{ab} = \ft14 \wtd T_{il}\, \wtd T_{j\ell}\, \bar\eta^a_{ij}\, 
\bar\eta^b_{k\ell}\,,\quad
P_{ab} = \ft14 \wtd T_{il}\, \wtd T_{j\ell}\, \eta^a_{ij}\, 
\bar\eta^b_{k\ell}\,.
\ee
After some manipulations, we can show that the matrices
$M$ and $N$ in (\ref{inversemet2}) are related to the matrix $P$ by
\be
M= (1+P \, P^t)^{1/2}\,,\qquad N= (1+ P^t\, P)^{1/2}\,,\label{mnp}
\ee
where $P^t$ denotes the transpose of $P$.  Thus the scalar fields in
the $S^3$ ansatz are parameterised purely by $P_{a b}$, which is in
the representation (Adjoint($SU(2)_L$), Adjoint($SU(2)_R$)) of the
$SU(2)_L \times SU(2)_R$ isometry group.  This exact result is in
accordance with the linearised scalar field analysis in \cite{het2}.

   We are now in a position to conjecture the form of the exact 
metric ansatz for the Pauli reduction of the bosonic string on any
group manifold $G$.  Namely, the proposal is that the inverse metric
is given by the same expression (\ref{inversemet2}), where now $R^a$ and 
$L^a$ are the Killing vectors associated with the right and left actions
of $G$ on the group manifold.  We again require also that (\ref{mnp})
hold, so that the scalar fields will be parameterised just by
$P_{ab}$, which is in the (Adjoint($G_L$),
Adjoint($G_R$)) representation of the group $G$.  This is in 
accordance with the linearised results in \cite{het2}, and indeed one can
check that our conjectured exact ansatz (\ref{inversemet2}) agrees after 
linearisation with the expression found in \cite{het2}.

\section{Superpotentials and Solutions}

   In this section, we return to the theme of DeWitt reductions of the
bosonic string, which we explored in section \ref{dewittred}.  We shall
now look for some explicit domain-wall solutions of the flow equations
in the DeWitt-reduced theories.

\subsection{Superpotentials for the truncated system}\label{superpotsec}

   We shall now examine the structure of the potential $V$ for the scalar
fields in $n$ dimensions in more detail.  To begin, it is convenient to
set to zero all the scalars described by the unimodular matrix $h_{\a\beta}$
(meaning that we take $h_{\a\beta}=\delta_{\a\beta}$), and also to set
to zero the scalars $\widetilde\chi_{\a\beta}$ coming from $\hat B_\2$.
The truncation of these fields can easily be seen to be consistent, and
the resulting potential $\widetilde V$ is given by
\be
\widetilde V = -\ft12 q\, e^{-\ft12 a\, \phi}\, 
(g^2\, e^{- \gamma\, \td\varphi}
- \ft13 m^2  e^{-3\gamma\, \td\varphi})\,.\label{tdv}
\ee

    The system of gravity coupled to the two remaining scalars $\phi$ 
and $\td\varphi$ is itself a consistent truncation of (\ref{fulllag}), 
and is described by the $n$-dimensional Lagrangian
\be
{\cal L} = R\, {*\oneone} - \ft12 {*d\phi}\wedge d\phi - 
\ft12{*d\td\varphi}\wedge d\td\varphi - \widetilde V\, {*\oneone}\,.
\label{scallag}
\ee
We can now seek a superpotential $W$, such that
\be
\widetilde V = \Big(\fft{\del W}{\del\phi}\Big)^2 +  
   \Big(\fft{\del W}{\del\td\varphi}\Big)^2
                 - \fft{(n-1)}{2(n-2)}\, W^2\,.\label{superpot}
\ee
It should be noted that the coefficient multiplying $W^2$ here must
necessarily have this specific value, if one is to be able to
formulate first-order equations that solve the equations of motion
following from (\ref{scallag}). (See, {\it e.g.} \cite{cgr,cs} and
references therein.)

   We can find three different choices for a superpotential $W$ such that
(\ref{superpot}) yields the truncated potential $\widetilde V$ 
given in (\ref{tdv}) (modulo inessential sign choices).
Each works only for a specific value of the dimension $q$ of the group
manifold $G$ on which the bosonic string was reduced; these dimensions
are $q=1$, 3 and 25, and so we shall denote the associated 
superpotentials by $W_1$, $W_3$ and $W_{25}$ respectively.  They are 
given by
\bea
W_1 &=& \fft{q}{2\sqrt6}\, e^{-\ft14 a\, \phi}\, 
(3g^2\, m^{-1}\, \, e^{\ft12\gamma\,\td\varphi} 
                 + m\, e^{-\ft32 \gamma\, \td\varphi})\,,\nn\\
W_3 &=&  \fft{q}{3\sqrt2}\, e^{-\ft14 a\, \phi}\, 
(-3g\, e^{-\ft12\gamma\,\td\varphi} 
                 + m\, e^{-\ft32 \gamma\, \td\varphi})\,,\label{superpots}\\
W_{25}&=&  \fft{q}{10\sqrt6}\, e^{-\ft14 a\, \phi}\, 
(5 g\, e^{-\ft12\gamma\,\td\varphi} 
                 - m^2\, g^{-1}\, e^{-\ft52 \gamma\, \td\varphi})\,.\nn
\eea
The constant $\gamma$ is given by (\ref{gammadef}) with $q=1$, 3 and 25 
respectively.  The case $q=1$ is vacuous, since there is no harmonic
3-form in a 1-dimensional group manifold.   We have no explanation, 
beyond the superficial calculational one, for why $q=3$ and $q=25$
should turn out to be only two cases where a superpotential leading
to first-order equations exists.

\subsection{Brane solutions from superpotentials}\label{superpotsolsec}

   Using these superpotentials, we can obtain first-order equations
for domain-wall configurations, and these equations can then be solved
explicitly.  To do this, we make the ansatz
\be
ds_n^2 = e^{2A}\, dx^\mu\, dx_\mu + dy^2\,,
\ee
where $A$, $\phi$ and $\td\varphi$ are assumed to depend only on the
transverse coordinate $y$.  From (\ref{scallag}), the  
equations of motion are given by
\bea
\phi'' + (n-1)\, A'\, \phi' &=& \fft{\del V}{\del\phi}\,,\nn\\
\td\varphi'' + (n-1)\, A'\, \td\varphi' &=& 
\fft{\del V}{\del\td\varphi}\,,\nn\\
A'' + (n-1)\, {A'}^2 &=& -\fft{V}{n-2}\,,\label{secondode}\\
\ft12 {\phi'}^2 + \ft12{{\td\varphi'}}{}^2 -(n-1)(n-2)\, {A'}^2 &=& V\,,\nn
\eea
where a prime denotes a derivative with respect to $y$.  It is easy to see
that these equations will be satisfied if the first-order equations
\be
\phi' = \sqrt2\,  \fft{\del W}{\del \phi}\,,\qquad
\td\varphi' = \sqrt2\,  \fft{\del W}{\del \td\varphi}\,,\qquad
A' = -\fft1{\sqrt2\, (n-2)}\, W\label{foeq}
\ee
hold. 

   We find the following solutions, where, without loss of generality, 
we have set $g=m$:

\noindent{$\bullet$ \underline{{\bf $q=25$}:}}

   Using $W=W_{25}$, we find that the first-order equations (\ref{foeq})
admit the domain-wall solution
\bea
ds_n^2 &=& e^{2A}\, (dx^\mu\, dx_\mu + H^{\ft12}\, dr^2)\,,\nn\\
e^{\ft12 a\phi} &=& e^{2A} = \Big(\fft{m^2\, r^2}{48}\Big)^{25/(n-2)}\,
  H^{5/(2(n-2))}\,,\qquad e^{\gamma\, \td\varphi} = \ft1{48} m^2\, r^2\, 
     H^{\ft12}\,,\label{dwsol25}
\eea
where $\gamma=\sqrt2/5$ and 
\be
H = 1 + \fft{2304}{m^4\, r^4}\,.
\ee
The solution (\ref{dwsol25}) can be lifted back to $(n+q)$ dimensions
using the DeWitt reduction ans\"atze of section \ref{dewittred},
yielding
\bea
d\hat s^2 &=& H^{-10/(n+23)}\, dx^\mu\, dx_\mu + H^{(n+3)/(2(n+23))}\, 
(dr^2 + r^2\, d\Omega_{25}^2)\,,\nn\\
e^{\ft12 \hat a\, \hat \phi} &=& H^{-10/(n+23)}\,,\qquad
\hat F_\3 = \ft16 m\, f_{\a\beta\gamma}\, \sigma^\a\wedge \sigma^\beta
\wedge \sigma^\gamma\,.\label{25sol}
\eea
where $d\Omega_{25}^2$ is the metric on the ``unit'' bi-invariant
25-dimensional group manifold $G$, scaled so that $R_{ij}= 24\delta_{ij}$.  
In terms of the left-invariant 1-forms $\sigma^\a$, it is given by
$d\Omega_{25}^2 = \ft1{48}\, g_{\a\beta}\, \sigma^\a\, \sigma^\beta$.  

   The lifted solution (\ref{25sol}) has a singularity at $r=0$, which
coincides with the horizon.  Note that the 26-dimensional space transverse
to the $(n-2)$-brane is nothing but the Ricci-flat cone over the bi-invariant
group manifold $G$.  Of course this transverse space also contributes its
own power-law curvature singularity at $r=0$, at the apex of the cone.
It is interesting to note, however, that in the string-frame metric,
related to the Einstein-frame metric by 
\be
d\hat s^2_{\rm string} = e^{-\ft12 \hat a\, \hat\phi}\, 
d\hat s^2_{\rm Einstein}\,,\label{stringeinst}
\ee
the solution takes the form
\be
d\hat s^2_{\rm string} = dx^\mu\, dx_\mu + H^{\ft12}\, (dr^2 + r^2\, 
d\Omega_{25}^2)\,.
\ee
This metric is completely non-singular, since $H\sim 1/r^4$ at small $r$. 
Thus the solution interpolates between $M_{n}\times G$ at small $r$, 
and $M_{n-1}$ times the Ricci-flat cone over $G$ at large $r$, where
$M_m$ denotes $m$-dimensional Minkowski spacetime.

\noindent{$\bullet$ \underline{{\bf $q=3$}:}}

   The superpotential $W_3$ in (\ref{superpots}) allows us to construct
a domain wall in $n$ dimensions, given by
\bea
ds^2 &=& e^{2A}\, (dx^\mu\, dx_\mu + H\, dr^2)\,,\nn\\
e^{\ft12 a\, \phi} &=& e^{2A}  = \Big(\fft{m^2\, r^2}{4}\Big)^{3/(n-2)}\, 
               H^{1/(n-2)}\,,\qquad
e^{\gamma\, \td\varphi} = \ft14 m^2\, r^2 \, H\,,
\eea
where $H=1 + 4/(m^2\, r^2)$.
Lifting back to $D=n+3$, and expressing the solution in the string frame,
we recover the standard $(D-5)$-brane, given by
\bea
d\hat s^2 &=& dx^\mu\, dx_\mu + H\, (dr^2 + d\Omega_3^2)\,,\nn\\
e^{\ft12 \hat a\, \hat \phi} &=& H^{-2/(n+1)}\,,\qquad 
F_\3 = m\, \Omega_\3\,,
\eea
where $d\Omega_3^2 = \ft14 \sigma_\a^2$ is the bi-invariant metric on 
the unit 3-sphere.

\subsection{Branes without superpotentials}

   We showed in section \ref{superpotsec} that in exceptional cases,
namely when the group manifold has dimension $q=3$ or $q=25$, the
scalar potential in the truncated $n$-dimensional theory
(\ref{scallag}) can be derived from a superpotential. In section
\ref{superpotsolsec} we made use of these superpotentials to construct
explicit domain-wall solutions, which we then lifted back to the
original $(n+q)$-dimensional bosonic string theory.  In this section,
we consider the generic situation when $q$ is equal to neither 3 nor
25, in which case we have no option but to solve the second-order
equations (\ref{secondode}) coming from (\ref{scallag}).  Since these
equations are quite involved, we shall resort to approximate methods
and numerical integration in order to study the form of the solutions.
Before doing this, however, we shall consider a simplified exact solution
in which the scalar field $\td\varphi$ is (consistently) set to zero.

   Setting $g=m$, we can set $\td\varphi=0$ in (\ref{secondode}), and then 
we can easily see that a solution is given by
\bea
ds_n^2 &=& e^{2A} \, dx^\mu\, dx_\mu + dy^2\,,\nn\\
e^{\ft12 a\phi} &=& e^{2A} = \fft{q\, m^2\, y^2}{3(n-2)^2}\,.
\label{case1}
\eea
Lifted to $(n+q)$ dimensions, this becomes
\bea
d\hat s^2 &=& \Big(\fft{q\, m^2\, y^2}{3(n-2)^2}\Big)^{(n-2)/(n+q-2)}\, 
\Big[ dx^\mu\, dx_\mu + m^{-2}\, g_{\a\beta}\, \sigma^\a\, \sigma^\beta +
      \fft{3(n-2)^2}{q\, m^2\, y^2}\, dy^2\Big]\,,\nn\\
e^{\ft12\hat a\, \hat \phi} &=&  
\Big(\fft{q\, m^2\, y^2}{3(n-2)^2}\Big)^{(n-2)/(n+q-2)}\,,\\
\hat G_\3 &=& \ft16 m\, f_{\a\beta\gamma}\, \sigma^\a\wedge 
\sigma^\beta\wedge \sigma^\gamma\,.\nn
\eea
In the string-frame metric, defined in (\ref{stringeinst}), the metric
becomes simply 
\be
d\hat s^2= dx^\mu\, dx_\mu + m^{-2}\, 
g_{\a\beta}\, \sigma^\a\, \sigma^\beta +
      \fft{3(n-2)^2}{q\, m^2\, y^2}\, dy^2 \,.
\ee
Note that $g_{\a\beta}\, \sigma^\a\, \sigma^\beta$ is just the 
bi-invariant metric on the group manifold $G$, with Ricci tensor given,
in orthonormal components, by $R_{ij} = \ft12 \delta_{ij}$.   In the 
string frame, the metric is totally regular; it is the direct sum of the 
metric on $n$-dimensional Minkowski spacetime and the 
bi-invariant metric on the group manifold $G$.

    Another special case corresponds to taking $m=0$.  The $n$-dimensional
solution is then given by  
\bea
ds_n^2 &=& e^{2A}\, (dx^\mu\, dx_\mu + dr^2)\,,\nn\\
e^{\ft12 a\, \phi} &=& e^{2A} = \Big( \fft{g^2\, r^2}{2(q-1)}\Big)^{q/(n-2)}
\,,\qquad e^{\gamma\, \td\varphi} = \fft{g^2\, r^2}{2(q-1)}\,,
\label{case2}
\eea
where $\gamma=\sqrt{2/q}$.  Lifted back to $(n+q)$ dimensions, it becomes
\bea
d\hat s^2 &=& dx^\mu\, dx_\mu + dr^2 + r^2\, d\Omega_q^2\,,\nn\\
\hat\phi &=& 0\,,\qquad \hat G_\3 =0\,,
\eea
where
\be
d\Omega_q^2 \equiv \fft1{2(q-1)}\, g_{\a\beta}\, \sigma^\a\, \sigma^\beta
\ee
is the ``unit'' bi-invariant metric on the group manifold $G$, whose
Ricci tensor satisfies $R_{ij} = (q-1)\, \delta_{ij}$.  Thus the
$(n+q)$-dimensional solution in this special case is just the
direct sum of the $(n-1)$-dimensional Minkowski metric and the Ricci-flat
metric on the cone over $G$.   

   We shall now demonstrate that there exists a more general brane solution
that interpolates between the solution (\ref{case1}) at short distance, and 
the solution (\ref{case2}) at large distance.  This more general solution
is the analogue of the explicit solutions that we found in section 
\ref{superpotsolsec} for the special cases $q=3$ and $q=25$.  Since the
second-order equations are quite difficult to solve explicitly, we
shall begin here by considering a Taylor expansion valid at short 
distances.  We find that the metric takes the form
\be
ds_n^2 = e^{2A}\, dx^\mu\, dx_\mu + dy^2\,,
\ee
with
\bea
e^{\ft12 a\phi} &=& e^{2A} = \fft{q\, m^2\, y^2}{3(n-2)^2}\,
    ( 1 + b_2\, z^2 + \cdots)\,,\nn\\
e^{-\gamma\, \varphi} &=& 1 + c\, z + c_2\, z^2 + \cdots
\eea
where
\bea 
z &=&y^{\ft{(n-2)[-q + \sqrt{q(q+24)}]}{2q}}\,,\nn\\
b_2 &=& -\ft{3q\,
c^2}{2[q + 24n-48 + \sqrt{q(q+24)}]}\,,\quad c_2=\ft{[-q-42 +
\sqrt{q(q+24)}]\,c^2}{2[-18 -q +\sqrt{q(q+24)}]}\,,
\eea
and $c$ is an integration constant.  We then apply numerical methods
using the above small-distance expansion to set initial data.  Lifting
back to $D=n+q$ dimensions, we find that for the solution to be free
from singularities in the string frame, the constant $c$ has to be
chosen so that $c\le 0$.  In the string frame, the metric interpolates
between $M_n\times G$ at small distance, and $M_{n-1}$ times the
Ricci-flat cone over $G$ at large distance.

\section{Discussion}

     A traditional approach to dimensional reduction, which became
popular in the 1980's, consisted of identifying a ``ground state'' of
the higher-dimensional theory that is (locally) a product of a
lower-dimensional spacetime $M$ and an internal (usually compact)
space $K$.  In other words, above each point $x$ in $M$ there was a
copy of $K$, called the fibre, whose metric varied as one moved about $M$.
One then considered small fluctuations around this
background.  If $K$ admitted isometries, then the fluctuations fell
into representations of the isometry group.  Usually, one focused
attention on the ``massless sector,'' which would comprise a finite
subset of the infinite towers of ``Kaluza-Klein modes.''  If the
internal space $K$ admitted isometries, then the massless sector would
be expected to include the Yang-Mills gauge bosons associated with the
isometry group of $K$. 

    At the linear level, one can of course always truncate out the
massive fields, since by definition there is no coupling between the
various fields.  At the full non-linear level, however, the question
of whether such a truncation may be consistently performed can be a
subtle one, since it might be that massless fields act as sources for
the massive fields that were originally set to zero.  There is no such
problem in the case of a circle or torus reduction, or for a DeWitt
group manifold reduction, since the ``massless sector'' is
characterised by its invariance under the simply-transitive action of 
the isometry group of the general fully non-linear ansatz.  Note that
in the DeWitt case the original ground state, with its bi-invariant 
metric on the group manifold $G$, has a larger isometry than that of
the general ansatz, since it admits the Killing vectors $R_a$ of $G_R$ as
well as the Killing vectors $L_a$ of $G_L$.\footnote{Recall that 
left-invariant vector fields $L_a$ generate right translations, 
while right-invariant vector fields $R_a$ generate left translations.}    

    For the case of a Pauli reduction, where $K$ is the coset space
$G/H$, there is no universal prescription for writing down a
reduction ansatz that would guarantee the full complement of gauge
bosons in a consistent reduction.  Indeed, for a Pauli reduction of
a generic theory it is guaranteed that no such ansatz exists; we have
discussed in previous sections some of the exceptional cases where a
consistent Pauli reduction is possible.

   More generally, one can consider a dimensional reduction in which
one makes no {\it a priori} assumptions about the nature of the
internal space $K$; it may not necessarily be related to a group
manifold or a coset space.  The induced metric $g_{mn}(x,y)$ on $K$
will vary from point to point on the base manifold, but only within a
finite-dimensional family, or modulus space ${\cal M}$, whose
coordinates $\phi$ define scalar fields $\phi(x)$ on spacetime.  The
modulus space may contain privileged ``ground-states'' which are
typically more symmetric than the general metric in ${\cal M}$. For
example, in the case of a DeWitt group-manifold reduction, the modulus
space ${\cal M}$ is the space of left-invariant metrics on the group
$G$, and the privileged metrics are the bi-invariant metrics, which
are unique up to a scale; the associated scalar field $\phi(x)$ is
called the ``breathing mode.''  In the case of Pauli coset reductions,
${\cal M}$ contains $G$-invariant ground-state metrics on $K=G/H$, but
it also includes metrics that are not invariant under $G$.  For a more
general reduction, such as a Calabi-Yau reduction, ${\cal M}$ will
contain no metrics invariant under any continuous group, and there may
not be any naturally-selected ``ground-state.''\footnote{Note however
that, as we shall discuss later, no Calabi-Yau reduction, of any
theory, can be expected to satisfy the strict requirement of
consistency that we are requiring in this paper.}

    An often-used starting point is a metric ansatz of the form
\bea
d\hat s^2 &=& g_{mn}(\phi(x),y)\, (dy^m - K^m_a(y)\, A^a_\mu(x)\, dx^\mu)
(dy^n - K^n_b(y)\, A^b_\nu(x)\, dx^\nu) \nn\\
&&+  W(\phi(x),y)^2\, g_{\mu\nu}(x)\, dx^\mu\,
dx^\nu\,,\label{genk}
\eea
where the $K_a^m(y)$ are the Killing vector fields (if any) of the
internal space $K$ in its ground state, $A^a_\mu(x)$ are the gauge
bosons associated with the group generated by these Killing vectors, and
$W(\phi(x),y)$ is a ``warp factor.''  The second term in (\ref{genk})
is the metric orthogonal to the fibres, and it is a conformal to
a fixed metric $g_{\mu\nu}$ on the base $M$.\footnote{Mathematically this
structure is an example of a ``conformal (pseudo)-Riemannian submersion.''
A Riemannian submersion in general has a metric of the form
\be
d\hat s^2 = g_{mn}(\phi(x),y)\, (dy^m - A^m_\mu(x,y) \,dx^\mu)
(dy^n - A^n_\nu(x,y)\, dx^\nu) 
+ g_{\mu\nu}(x)\, dx^\mu\,
dx^\nu\,.\label{on}
\ee
The ansatz (\ref{genk}) differs from this in that the cross terms take
a particular factorised form and in general a $y$-dependent conformal
factor is inserted in front of the second term.}
If one wants the
lower-dimensional metric $g_{\mu\nu}(x)$ to describe gravity in the
Einstein conformal frame (\ie with a canonical $\sqrt{-g}\, R$
Einstein-Hilbert action), then the warp factor should be chosen to be
\be
W= \Big( \fft{\sqrt{\det(g_{mn}(\phi(x),y))}}{
        \sqrt{\det(g_{mn}(y))}}   \Big)^{-\ft1{n-2}}\,,
\label{warpfactor}
\ee
where the lower-dimensional spacetime has dimension $n$, and   
$g_{mn}(y)$ is any $x$-independent metric on $K$. This ``fiducial''
metric on $K$ is needed in order to ensure that $W$ is a scalar
with respect to coordinate transformations of the $y^m$. A convenient
choice is to take the fiducial metric to be the ground-state metric.

    It is interesting to compare the ansatz (\ref{genk}) with the
exact answer in known cases. First, consider a DeWitt reduction, 
whose metric ansatz is
\be
d\hat s^2 = g_{ab}(x)\, (\lambda^a - A^a)(\lambda^b-A^b) + 
           (\det g_{ab}(x))^{-\ft1{n-2}} \, 
g_{\mu\nu}(x)\, dx^\mu\, dx^\nu\,,\label{dwx}
\ee
where the conformal factor in the second term has been chosen 
so that the lower-dimensional metric is in the Einstein conformal frame.
Since the left-invariant 1-forms can be written as $\lambda^a =
\lambda^a_m(y)\, dy^m$, we can  rewrite (\ref{dwx}) in the 
form of (\ref{genk}), with 
\be
g_{mn}(x,y) = g_{ab}(x)\, \lambda^a_m(y)\, \lambda^b_n(y)\,,\qquad
   K^m_a(y) = L^m_a\,,\qquad W^2 = (\det g_{ab}(x))^{-\ft1{n-2}}\,,
\ee
where $L^m_a$ are the vector fields dual to $\lambda^a$; \ie $L^m_a\,
\lambda^b_m = \delta^b_a$.  In other words $L^m_a$ are the Killing
vectors of {\sl right-translations} of the bi-invariant metric on $G$.
If we compare with the ansatz (\ref{genk}), we see that the Killing
vectors $K^m_a$ in this case coincide with half of the total
isometries of the background, \ie of the bi-invariant metric on $G$,
and they are not Killing vectors of the generic deformation described
by the ansatz (\ref{dwx}).  In fact in this case, one might have hoped
to obtain an ansatz that included the gauge bosons associated both
with $G_R$ and $G_L$, but the DeWitt ansatz includes only those
associated with $G_R$.  As we discussed in section 6.2, while a
consistent reduction including the gauge bosons of $G_L\times G_R$
definitely cannot be consistent if one begins with a generic
higher-dimensional theory, it is believed that for the special case of
the $D$-dimensional bosonic string such a consistent reduction is
possible.  Of course this is really a Pauli reduction on $G$ viewed as
the coset $(G\times G)/G$, rather than a DeWitt reduction on
the group manifold $G$.  Note, incidentally, that the ``warp factor''
$W^2$ in a DeWitt reduction is {\sl independent} of the coordinates
$y^m$ of the internal space.

    We can also make a comparison between (\ref{genk}) and the 
metric ansatz given in (\ref{s2pauli}) for the Pauli reduction of 
Einstein-Maxwell-dilaton gravity on $S^2$.  A convenient way to do
this is by choosing $y_1 = \mu_1$ and $y_2=\mu_2$ as independent
coordinates on $S^2$, with $\mu_3= \sqrt{1-y_1^2 -y_2^2}$.  It is
straightforward to see from the definition of $D\mu^i$ that we shall
have
\be
D\mu^m = dy^m - K^m_i\, A^i\,,
\ee
where $K_i = -\epsilon_{ijk} \mu^j\, (\del/\del\mu^k)$ are the Killing
vectors on the ground-state 2-sphere.  In terms of the unconstrained
coordinates $y^m$ ($m=1,2$) we have
\be
K_1 = \mu_3\, \fft{\del}{\del y_2}\,,\quad 
K_2 = - \mu_3\, \fft{\del}{\del y_1}\,,\quad 
K_3 = y_2 \, \fft{\del}{\del y_1} - y_1\, \fft{\del}{\del y_2}\,.
\ee
Thus we can express the Pauli metric reduction (\ref{s2pauli}) as
\be
ds_D^2 = g_{mn}\, (dy^m - K^m_i\, A^i)(dy^n- K_j^n\, A^j) +
 (Y\, \Delta)^{\ft1{D-2}}\, ds_{D-2}^2\,,\label{paulixx}
\ee
where
\be
g_{mn} = (T^{-1})_{mn} - \mu_3^{-1}\, (T^{-1})_{m3}\, y^n - 
  \mu_3^{-1}\, (T^{-1})_{n3}\, y^m + \mu_3^{-2}\, (T^{-1})_{33}\,
y^m\, y^n\,.
\ee
A straightforward calculation shows that $\det (g_{mn})=
\Delta/(\mu_3^2\, \det(T_{ij})) = \Delta/ (\mu_3^2\, Y)$, where as
usual $\Delta=T_{ij}\, \mu^i\, \mu^j$.  Note that for the ``fiducial
metric'' corresponding to $T_{ij}=\delta_{ij}$, we have $\det(\bar
g_{mn})=\mu_3^{-2}$.   Substituting into (\ref{warpfactor}), we find
that the warp factor is given by
\be
W^2 = (Y\, \Delta)^{\ft1{D-2}}\,,
\ee
which is indeed the factor appearing in the Pauli reduction ansatz
(\ref{paulixx}).  Thus we have confirmed that the $S^2$ Pauli metric
reduction ansatz is indeed of the general form given in (\ref{genk}).
One can straightforwardly show that all the other known Pauli
reduction examples, such as the $S^3$ reduction of the bosonic string,
also fit the general form of the reduction ansatz (\ref{genk}).  Note
that in the Pauli coset reduction, by contrast to the DeWitt
group-manifold reduction, the warp factor $W^2$ depends on the
coordinates $y^m$ of the internal space as well as on the coordinates
$x^\mu$ on the base.  For the reader's convenience, the gauge group
structures of Kaluza, DeWitt and Pauli reductions are summarised in
Table 1 below.

\bigskip\bigskip
\centerline{
\begin{tabular}{|c|c|c|c|}\hline
  & Kaluza & DeWitt & Pauli \\ \hline\hline 
Internal Space, $K$ & $U(1)$  & $G$ & $G/H$  \\ \hline 
Gauge Bosons & $U(1)$ & $G_R$  & $G_R$ \\ \hline
Generic Isometries & $U(1)$ & $G_L$ & None \\ \hline 
Ground-state Isometries& $U(1)$ & $G_L\times G_R$ & $G_R$ \\ \hline 
\end{tabular}}
\bigskip

\noindent{\bf Table 1:} 
The gauge bosons, isometries for generic scalar configurations, and
isometries for the most symmetric ``ground-state'' configuration for
the various dimensional reduction schemes.
\bigskip

   While (\ref{genk}) is a general proposal for the metric 
reduction ansatz, which appears to work in all known cases where a
consistent reduction is possible, it is very far from providing a complete
ansatz because it says nothing about the reduction of any matter
fields.  As we have already remarked, consistent Pauli reductions can
never work starting from pure higher-dimensional gravity, and it is
only in very exceptional cases that they can work even when matter is
included.  To see this more clearly, it is instructive to see what
happens when one substitutes the general metric reduction ansatz
(\ref{genk}) into the higher-dimensional Einstein equations.  In
practice this is rather complicated if one retains the scalar fields,
but in fact the relevant points can be made adequately if we set the
scalars to zero; \ie we take $g_{mn}(\phi(x),y)$ to be the
ground-state metric $g_{mn}(y)$ on $K$, and we take $W(\phi(x),y)=1$.  
If one then substitutes the ansatz into the lower-dimensional
components of the higher-dimensional equation $\hat R_{AB}=0$, one
finds (see, for example, \cite{homapo})
\be
R_{\mu\nu} - \ft12 R\, g_{\mu\nu} = \ft12 Y_{ab} \, 
(F^a_{\mu\rho}\, F^{b\, \rho}_{\nu} - \ft14 F^a_{\rho\sigma}\, F^{b
\, \rho\sigma}\, g_{\mu\nu}) + \ft12 R(K)\, g_{\mu\nu}\,,\label{inconeinst}
\ee
where $R_{\mu\nu}$ and $R$ are the lower-dimensional ($y$-independent)
Ricci tensor and scalar, 
and $R(K)$ is the Ricci scalar of the
ground-state metric on the internal space $K$.  In this example, 
where we are reducing pure Einstein gravity, 
$Y_{ab}= g_{mn}\, K^m_a\, K^n_b$.
The general problem of
inconsistency of the reduction can be seen from the fact that the
product of Killing vectors in $Y_{ab}$  is in
general $y$-dependent, and so (\ref{inconeinst}) is not a consistent
lower-dimensional Einstein equation.  There is no problem for Kaluza
or DeWitt reductions, since the product of Killing vectors in 
$Y_{ab}$ {\sl is} then $y$-independent.
For Pauli reductions on cosets, $Y_{ab}$ depends on $y$, and
consistency is only achievable, if at all, when additional matter is
included in the higher-dimensional theory whose reduction ansatz
contributes additional terms in (\ref{inconeinst}) that ``conspire''
to produce a $y$-independent prefactor $Y_{ab}$ for the uncontracted
``stress tensor'' $(F^a_{\mu\rho}\, F^{b\, \rho}_{\nu} - \ft14
F^a_{\rho\sigma}\, F^{b \, \rho\sigma}\, g_{\mu\nu})$ in the analogue
of (\ref{inconeinst}).  For example,
in the $S^7$ reduction of eleven-dimensional supergravity, one gets
additional contributions from the reduction ansatz for the 4-form,
leading to 
\be
Y_{ab}=  g_{mn}\, K^m_a\, K^n_b +\fft1{2m^2}\, g^{mn}\, g_{pq}\, 
\nabla_m\, K^p_a\, \nabla_n\, K^q_b\,,\label{combo}
\ee
where $R_{mn}= (q-1)  m^2\, g_{mn}$, with $q=7$.
 The  combination in (\ref{combo}) is,
in fact, $y$-independent on any sphere $S^q$.
In  \cite{dunipowa},  a discussion of
the $S^7$ reduction was given. A more general analysis of related
consistency issues appears in 
\cite{homapo}, including a discussion of why the restoration
of the scalars, which of course would be necessary for a full
discussion of consistency, could not resolve the above inconsistency 
in this particular sector. Thus finding, by virtue of some
``conspiracy'' involving additional matter contributions, that 
the tensor $Y_{ab}$ in the lower-dimensional 
components of the higher-dimensional Einstein equation is
 $y$-independent is a {\it
 necessary}, but not {\it sufficient}, condition for having a 
consistent reduction ansatz.  In the light of these discussions,
it seems very unlikely that consistency could ever be achieved
for ground state manifolds $K$ with non-transitively acting
isometry groups, regardless of of the choice of additional matter.

   It is worth remarking that in the known cases of consistent
Pauli reductions the difficulty in finding the reduction ansatz for
the form fields is enormously greater than that for the metric
reduction ansatz.  (A striking example is provided by the $S^7$
reduction of eleven-dimensional supergravity, for which a proof of
consistency is presented in \cite{dwinic}.  Although explicit formulae
for the metric reduction are given, there is no complete explicit
result for the reduction of the 4-form field.)

    In this paper, we have concentrated on the discussion of Kaluza
and DeWitt group-manifold reductions, where consistency is guaranteed,
and Pauli coset reductions, where consistency is achieved only in
certain exceptional cases.  Reduction ans\"atze are sometimes
considered in more general cases where the internal space $K$ is
inhomogeneous, admitting either an intransitively-acting group of
isometries, or no continuous isometries at all.  In the rare examples
of consistent Pauli reductions, it is only as a consequence of very
remarkable ``conspiracies'' between the properties of the original
higher-dimensional theory, and the specific reduction manifold, that the
consistency is achievable at all.  It is evident that for an
inhomogeneous internal space $K$, the likelihood of any such analogous
conspiracies arising is rather remote.  Thus it would seem to be
highly unlikely that any kind of non-trivial consistent reduction is
possible for any inhomogeneous internal space.  For example, there
would appear to be no reason to expect that a consistent reduction on
a Calabi-Yau manifold $K$ would ever be consistent, if one tried to
retain the set of massless scalars corresponding to the moduli of $K$,
while setting the massive modes to zero.  This is because one can
expect that there will be non-linear couplings in the full untruncated
theory in which powers of the massless scalars act as sources for the
massive scalars that one is trying to set to zero.  Thus, while we are
not in a position to offer a cast-iron proof of the inconsistency in
such a case, we can assert that the ``burden of proof'' lies with
those who would claim that such reductions are
consistent.\footnote{This is quite a different matter from the
question of whether one can, to a very good approximation, neglect the
effect of ignoring the massive modes and their non-linear massless
source-terms in a discussion of low-energy effective physics in
four-dimensional string compactifications.  Here, we are addressing a
mathematical issue of exact embeddability, not a physical question of
approximate decoupling of massive modes.}

\section*{Acknowledgments}

M.C. (on sabbatical leave from the University of Pennsylvania) is
supported in part by DOE grant DE-FG02-95ER40893, NATO linkage grant
No. 97061, National Science Foundation Grant No. INT02-03585 and the
Fay. R. and Eugene L.  Langberg Chair.  M.C. would like to thank the
Univ. of Texas A\&M for hospitality during the initial stages of this
work and the Institute for Advanced Study, Princeton, for hospitality
and support during the course of this work.  H.L and C.N.P. are
supported in part by DOE grant DE-FG03-95ER40917.  G.W.G. would like
to than the Director and staff of the George P. and Cynthia
W. Mitchell Institute for Fundamental Physics for their hospitality
during the course of this work.  C.N.P. would like to thank the
Cambridge Relativity and Gravitation Group for hospitality during the
completion of this work.

\appendix
\section{DeWitt = Pauli $\circ $ DeWitt: Reductive Coset Reductions}

   In this appendix, we derive explicit expressions for the DeWitt
reduction on the group manifold $G$ can be viewed as an initial DeWitt
reduction on the subgroup $H$, followed by a Pauli reduction on the
coset $G/H$, in cases where the coset is reductive.

    Using the appropriate definitions given in sections \ref{gcase} and
\ref{ghcase}, we have
\bea
\rho^a\, T_a &=& dh\, h^{-1} + h\, dk\, k^{-1}\, h^{-1}\nn\\
&=& (R^\a +\omega^\a)\, T_\a + \rho^i\, T_i\,,\label{app1}
\eea
where the $R^\a$ are right-invariant 1-forms on $H$, 
\be
R^\a\, T_\a = dh\, h^{-1}\,,
\ee
and $h\, dk\, k^{-1}\, h^{-1} = \omega^\a\, T_\a + \rho^i\, T_i$.
We also have
\bea
\lambda^a\, T_a &=& k^{-1}\, h^{-1}\, dh\, k + k^{-1}\, dk = 
 k^{-1}\, (\Lambda^\a\, T_\a)\, k + k^{-1}\, dk\nn\\
&=& T_a\, U^a{}_\a\, (R^\a + \omega^\a) + T_a\, U^a{}_i\, \rho^i\nn\\
&=& T_a\, U^a{}_\a\, W_\beta{}^\a\, (\Lambda^\beta + W^\beta{}_\gamma\, 
\omega^\gamma) + T_a\, U^a{}_i\, \rho^i\,,\label{lam7}
\eea
where $\Lambda^\a\, T_\a = h^{-1}\, dh$ defines the left-invariant
1-forms on $H$, and $W^\a{}_\beta$ is defined for $H$, in a manner
analogous to the definition of $U^a{}_b$ for $G$, by
\be
T_\a\, W^\a{}_\beta = h^{-1}\, T_\beta\, h\,.
\ee
The next task is to show that the various terms in the final line
of (\ref{lam7}) are all independent of the $H$ subgroup coordinates
$\tau_\a$ (aside from their appearance in the left-invariant 1-forms
$\Lambda^\a$ themselves).

    First, we note that
\bea
T_a\, U^a{}_\a\, W_\beta{}^\a &=& V^{-1}\,T_\a\, V\, W_\beta{}^\a 
=k^{-1}\, h^{-1}\, T_\a\, h\, k\, W_\beta{}^\a = k^{-1}\, T_\gamma\, k\, 
W^\gamma{}_\a \, W_\beta{}^\a\nn\\
&=& k^{-1}\, T_\beta\, k\,,
\eea
thus showing that $U^a{}_\a\, W_\beta{}^\a$ is independent of $\tau_\a$.

     Next, we note from (\ref{app1}) that
\bea
dk\, k^{-1} &=& \omega^\a\, h^{-1}\, T_\a\, h + \rho^i\, h^{-1}\, T_i\, 
h\nn\\
&=& \omega^\a\, W^\beta{}_\a\, T_\beta + \rho^i\, h^{-1}\, T_i\, h\,.
\eea
If the coset is reductive, meaning in particular that $[H,K]=K$, 
we see that $h^{-1}\, T_i\, h = B_i{}^j\, T_j$ for some ($\tau_\a$-dependent)
matrix $B_i{}^j$.   Since the assumed reductivity means that there
are no $T_\a$ terms coming from $h^{-1}\, T_i\, h$, it follows that
the manifest $\tau_\a$-independence of $dk\, k^{-1}$ implies that 
we can independently deduce that $\omega^\a\, W^\beta{}_\a$ and 
$\rho^i\, h^{-1}\, T_i\,  h$ are independent of the coordinates $\tau_\a$.
Finally, from
\bea
T_a\, U^a{}_i\, \rho^i &=& V^{-1}\, T_i\, V\, \rho^i =
k^{-1}\, h^{-1}\, T_i\, h\, k\, \rho^i\nn\\
&=& k^{-1}\, (\rho^i\, h^{-1}\, T_i\, h)\, k\,,
\eea
the $\tau_\a$-independence of $\rho^i\, h^{-1}\, T_i\, h$ allows us
to deduce that $U^a{}_i\, \rho^i$ is independent of $\tau_\a$.

   It is now apparent that with $\lambda^a$ given by the final line of
(\ref{lam7}), \ie
\be
\lambda^a =  U^a{}_\a\, W_\beta{}^\a\, (\Lambda^\beta + W^\beta{}_\gamma\, 
\omega^\gamma) + U^a{}_i\, \rho^i\,,
\ee
we can re-express the standard DeWitt metric reduction (\ref{dewitt7}),
after completing the square on the $\Lambda^\a$ terms, as
\bea
d\hat s^2 &=& e^{2\a\,\varphi}\, ds^2 + 
e^{2\beta\, \varphi}\, \Delta_{\a\beta}\, 
(h^\a + \Delta^{\a\gamma}\, T_{ac}\, M^a{}_\gamma\, h^c)
(h^\beta + \Delta^{\beta\delta}\, T_{bd}\, M^b{}_\delta\, h^d)
\nn\\
&& + e^{2\beta\, \varphi}\, (T_{ab} - \Delta^{\a\beta}\, T_{ac}\, 
T^{bd}\,M^c{}_\a\, M^d{}_\beta) \, h^a\, h^b \,,\label{hred}
\eea
where we have defined 
\bea
h^a &\equiv& U^a{}_i\, (\rho^i - U_b{}^i\, A^b)\,,\qquad
h^\a \equiv \Lambda^\a + W^\a{}_\beta\, (\omega^\beta - 
U_a{}^\gamma\, A^a)\,,\nn\\
M^a{}_\beta &\equiv& U^a{}_\gamma\, W_\beta{}^\gamma\,,\qquad
\Delta_{\a\beta} \equiv T_{ab}\, M^a{}_\a\, M^b{}_\beta\,,
\eea
and $\Delta^{\a\beta}$ is the inverse of $\Delta_{\a\beta}$; \ie 
$\Delta^{\a\beta}\, \Delta_{\beta\gamma} = \delta^\a_\gamma$.  
Since all the quantities in (\ref{hred}) are independent of
$\tau_\a$, except for the left-invariant 1-forms $\Lambda^\a$ themselves,
the expression (\ref{hred}) can now be viewed as a metric 
ansatz for a standard DeWitt reduction on the group manifold $H$ with
left-invariant 1-forms $\Lambda^\a$.

    After performing this DeWitt reduction on $H$, we obtain a theory
which can then undergo a consistent Pauli reduction on the 
coset space $G/H$.

\section{Further Examples of Coset Reductions}

\subsection{The case of $(G\times G)/G$}

   Any group may be considered as a coset of the product of the group
with itself with respect to the diagonal subgroup. This fact
plays an important role in string theory and so we will describe here
some of the relevant geometry.  For clarity, and consistency with our
previous notation  we take  $G\times G$ as pairs
$g=(g_1, g_2)$, with $g_1, g_2 \in G$, with the usual multiplication rule,
and the diagonal subgroup $H=G_{\rm diag}$,  
as pairs $h=(u,u)$.  We can, at least in the neighbourhood
of the identity,  write uniquely $G\times G =H K$, 
\ben
(g_1,g_2)= (u,u) \, (v, v^{-1})
\een
with $h=(u,u)$ and $k= (v, v^{-1})$. It follows that $g_1= u\, v$,
$g_2=u\, v^{-1}$\,,
and thus
\ben
w=v^2= g_2^{-1}\,  g_1 \qquad u= g_1 \, v^{-1}\,.
\een
Thus as long as the square root $( g_2^{-1}\,  g_1)^{ 1 \over 2} $
exists and is unique, we have a unique decomposition. This will be true
as long as the exponential map is onto.

   Defining $D\, g \equiv d\, g - g\, A$, where the gauge potentials 
$A$ of $G\times G$ are expressed as $A=(A_1,A_2)$, one has
\bea
g^{-1} D\, g&=& k^{-1}\, (h^{-1}\, dh + dk\, k^{-1} - k\, A\, k^{-1})
\, k \label{form1}\\
 &=& \!\! \! \!
(v, v^{-1} )^{-1} \Bigl ( (u^{-1}  du, u^{-1}\,  du ) + ( dv
\, v^{-1}, - v^{-1} \, dv) +(v\, A_1\, v^{-1}, v^{-1}\, A_2\, v) 
\Bigl ) (v, v^{-1})\,,\nn
\eea
and thus
\ben
h^{-1} \, dh = (u^{-1} \, du, u^{-1}\, du)
\een
is along the fibre. We now need to project the remaining terms in
(\ref{form1}) parallel and perpendicular to the orbits of the diagonal
subgroup $H$, as discussed in section \ref{ghcase}. 
To do this we need a metric on the group $G\times G$,
and we take the Killing metric which of course coincides with the
product of the Killing metrics on the two factors.  Thus elements of
the Lie algebra in $(\frak{h}, \frak{h})$ are parallel, whilst
elements in $(\frak{h}, -\frak{h})$ are perpendicular.  We therefore
have
\bea
&&(dk\, k^{-1} - k\, A\, k^{-1})_\|  \\
&& =\ft12 ( dv \, v^{-1} - v^{-1}\,  dv + v\, A_1\, v^{-1} + v^{-1}\, A_2\, v, 
dv\, v^{-1} -v^{-1} \, dv + v\, A_1\, v^{-1} + v^{-1}\, A_2\, v)\,,\nn
\eea
and
\ben
(dk\, k^{-1})_\perp =\ft12(dv\,  v^{-1} + v^{-1} \, dv
+ v\, A_1\, v^{-1} - v^{-1}\, A_2\, v
, -dv \, v^{-1} -
v^{-1} \, dv   - v\, A_1\, v^{-1} + v^{-1}\, A_2\, v)\,.
\een

    The DeWitt metric reduction ansatz for this case can then be
written in the standard way.  If, for simplicity, we omit the scalars
(which would not, in general, be a consistent thing to do), the metric
is then given by
\ben
  ds^2 = {\rm Tr}(u^{-1} \, du + \omega 
+ \ft12 v\, A_1\, v^{-1} + \ft12 v^{-1}\, A_2\, v)^2  
+ \ft14 {\rm Tr }
  (dv \, v^{-1} + v^{-1} \, dv + v\, A_1\, v^{-1} - v^{-1}\, A_2\, v)^2\,,
 \een
 with
\ben
\omega \equiv  \ft12(dv \, v^{-1} -v^{-1} \, dv)\,.
\een
The round metric induced on the coset $(G\times G)/G$ itself 
is thus
\ben
ds^2_{\rm Base} =\ft12 {\rm Tr }(
dv \, v ^{-1} + v^{-1} \, dv )^2\,.
\een
Remarkably, this may be expressed entirely in terms of $ w=v^2$, as
\ben
ds ^2_{\rm Base}  = {1 \over 4} {\rm Tr } 
\bigl ( w^{-1} \, dw \, w^{-1} \, dw \bigr ).
\label{identiity}
\een
We have  arrived at the bi-invariant metric on $H$, as expected, even
though $dv\, v^{-1} + v^{-1} \, dv$ is the sum of left-invariant
and right-invariant one forms, and thus has a well defined
transformation only under the adjoint action of $H$ on $H$, \ie
$v\longrightarrow U^{-1}\, v\, U$.

    It is instructive to consider the special case of $H=SU(2)$.
We set
\ben
v= \pmatrix { t+iz & x+iy &\cr -x+i y & t-iz &\cr }\,,
\een
and
 \ben
w= \pmatrix { \mu_0 +i\mu _3  & \mu_1 +i\mu _2  &\cr -\mu_1+i \mu _2
 & \mu_0 -i\mu _3 &\cr } \,.
\een
One has
\ben
\mu_0 = t^2 -x^2 -y^2 -z^2, \qquad \mu_i = 2 t x_i\,,
\een
with $i=1,2,3$. The inverses are
\ben
t= \pm {1\over \sqrt  2}\sqrt { 1+\mu_0}, \qquad x_i = {\mu _i \over
  \sqrt{2 (1+\mu_0 ) }}\,.
\een

\subsection{The case of $S^n = SO(n+1)/SO(n)$}

   As a further example, we shall work through in detail
the case of $S^n = SO(n+1)/SO(n)$.  Since the inclusion
of the gauge fields is straightforward, but adds little to the
geometrical discussion that we wish to give here, we shall omit
them in what follows.

   We express an arbitrary element $g$  of $SO(n+1)$ as $g=h\, g$, with
$h\in SO(n)$.  It is convenient to write the generators of $SO(n+1)$ 
as $\Sigma_{AB}= -\Sigma_{BA}$, and to decompose the fundamental
index as $A=(0,i)$, so that the generators of the $SO(n)$ subgroup
are $\Sigma_{ij}$.  In a matrix representation
\ben
h= \exp \thinspace  {\pmatrix { \L _{ij} & 0   \cr  0 & 0 \cr}}\,,
\een
and
\ben
k= \exp \thinspace {\pmatrix { 1 & {\bf b} \cr -{\bf b}^t & 0 \cr }}=
\pmatrix { 1-{{\bf x} {\bf x}^t \over 1+ x_0} & {\bf x}
 \cr -{\bf x}^t & x_0 \cr },
\een
where
 ${\bf x}={\bf  b} \, {\sin b \over b}$ with $b=|{\bf b}|$, $x_0=
\sqrt{1-{\bf x}^2 }$ and $x^2 = {\bf x}^t \, {\bf x}$. Note
that if we take $x_0 >0$, we cover one half of the sphere, whilst if  we take
$x_0<0$ we cover the other half.
The left-invariant forms on the coset are
\ben
k^{-1} dk= {\pmatrix {
  { {\bf x} d{\bf x} ^t- d{\bf x} {\bf x} ^t\over
1+x_0 } & d{\bf x} -{ {\bf x} d x_0 \over 1+ x_0} \cr -d{\bf x}^t  +
{ {\bf x} ^t d x_ 0 \over 1+ x_0 } & 0 \cr }}\,,
\een
while the right-invariant one forms are
\ben 
dk k^{-1} = {\pmatrix {
  { d{\bf x} {\bf x} ^t-  {\bf x} d  {\bf x}^t  \over 1+x_0 }
 & d{\bf x} -{ {\bf x} d x_0 \over 1+ x_0} \cr -d{\bf x}^t  +
{ {\bf x} ^t d x_ 0 \over 1+ x_0 } & 0 \cr }}\,.
\een

   We denote the left-invariant 1-forms on $SO(n)$ by
$\bigl (h^{-1} dh \bigr ) _{ij}= \sigma _{ij}$.  The projection of 
$dk\, k^{-1}$ that is parallel to the $SO(n)$ fibres is given by
\ben
\bigl (dk\, k^{-1}   \bigr )_{ij}\equiv \omega _{ij} 
= { x_i d x_j - x_j d x_i
\over 1 + x_0 } = -\bigl (k^{-1} \, dk \bigr )_{ij}\,.
\een
The projection perpendicular to the fibres is given by
\ben
K_i\equiv  \bigl ( dk\, k^{-1} \bigr )_{0i}= 
dx_i -{ x_i dx_o \over 1+ x_0}= -\bigl
( k^{-1} \, dk \bigr )_{0i}\,.
\een

   The bi-invariant metric on $SO(n+1)$ is given by
\ben
ds^2 = - \ft12 {\rm Tr } \bigl (g^{-1} dg \bigr ) ^2= -\ft12
{\rm Tr} \bigl ( h^{-1} dh + dk k^{-1}  \bigr ) ^2 .
\een
Using the formulae above one gets
\ben
ds^2 = \ft12 \bigl (\sigma _{ij} +\omega_{ij}  \bigr )^2  +K_i\, K_i\,,
\een
with
\ben
K_i\, K_i = dx_0^2 + dx _i^2\, ,
\een
which is the metric on the $S^n$ base of this principal
$SO(n)$ bundle (\ie the bundle of orthonormal frames over $S^n$).
The 1-forms $\sigma _{ij}$ span the $SO(n)$ fibres
and the 1-forms $\omega_{ij}$ give the horizontal connection
with respect to the Killing metric. 

    From the point of view of dimensional reduction, both the bundle
(\ie $SO(n+1)$) and the base (\ie $S^n$) are Einstein
spaces. Dimensional reduction of the Einstein action on $G=SO(n+1)$
gives $SO(n+1)$ Einstein-Yang-Mills theory coupled to scalars in the
symmetric tensor representation of the adjoint. The  ``breathing
mode'' corresponds to the determinant of this symmetric tensor, which
is an $SO(n+1)$ singlet. Because of the cosmological term in the bundle,
this singlet scalar will have a potential.  In an Einstein solution given by
the Killing metric, one of the  scalars are excited.   Yang-Mills 
fields are present in this background,  whose
potentials are given precisely by the $\frak {s} \frak {o}
(n)$-valued one-forms $\omega_{ij}$.


\begin{thebibliography}{99}

\bm{kaluza1} T. Kaluza, {\it On the problem of unity in physics},
Sitzunber. Preuss. Akad. Wiss. Berlin. Math. Phys. {\bf K1}, 966 (1921).

\bm{klein1} O. Klein, {\it Quantum theory and 5-dimensional theory of
relativity}, Z. Phys. {\bf 37}, 895 (1926).

\bm{jordan} P. Jordan, {\it Extension of projective relativity}, 
Ann. der Phys., Lpz. {\bf 1}, 219 (1947).

\bm{thiry} Y. Thiry, {\it On the regularity of gravitational and 
electromagnetic fields in unitary theories. I \& II}, C.R. Acad. Sci. Paris
{\bf 226}, 216 and 1881 (1948).

\bm{pauli} W. Pauli, Wissenschaftlichter Briefwechsel, Vol. IV, Part II
(1999), Springer-Verlag, edited by K.V. Meyenn.

\bm{strau} N. Straumann, {\it On Pauli's invention of non-Abelian
Kaluza-Klein theory in 1953}, gr-qc/0012054.

\bm{oraf} L. O'Raifeartaigh and N. Straumann, {\it Early history of
gauge theories and Kaluza-Klein theories, with a glance at recent
developments}, hep-ph/9810524.

\bm{klein2} O. Klein, in {\it New theories in physics}, International 
Institute of Intellectual Cooperation, Paris, 1939. 

\bm{gross} D.J. Gross, {\it Oscar Klein and gauge theory}, 
hep-th/9411233.

\bm{dewitt} B.S. DeWitt, in {\it Relativity, groups and topology},
Les Houches 1963 (Gordon and Breach, 1964).

\bm{het1} M.J. Duff, B.E.W. Nilsson and C.N. Pope,
{\it Kaluza-Klein approach to the heterotic string},
Phys. Lett. {\bf B163}, 343 (1985).

\bm{kerner} R. Kerner, {\it Generalisation of the Kaluza-Klein theory
for an arbitrary non-Abelian gauge group}, Ann. Inst. H. Poincar\'e
{\bf 9}, 143 (1968).

\bm{chofre} Y.M. Cho and P.G.O. Freund, {\it Non-Abelian gauge fields as
Nambu-Goldstone fields}, Phys. Rev. {\bf D12}, 1711 (1975).

\bm{schsch} J. Scherk and J.H. Schwarz, {\it How to get masses from 
extra dimensions}, Nucl. Phys. {\bf B153}, 61 (1979).


\bm{hawking} S.W. Hawking, {\it On the rotation of the universe}, 
Mon. Not. Roy. Astr. Soc. {\bf 142}, 129 (1969).

\bm{sneddon} G.E. Sneddon, {\it Hamiltonian cosmology: a further 
investigation}, J. Phys. Math. Gen. {\bf 9}, 229 (1976).

\bm{dwinic} B. de Wit and H. Nicolai, {\it The consistency of the 
$S^7$ truncation In D = 11 Supergravity},
Nucl. Phys. {\bf B281}, 211 (1987).

\bm{nasvamvan} H. Nastase, D. Vaman and P. van Nieuwenhuizen,
{\it Consistency of the AdS$_7\times S^4$ reduction and the origin of  
self-duality in odd dimensions},
Nucl. Phys.  {\bf B581}, 179 (2000), hep-th/9911238.

\bm{s5red}
M. Cveti\v c, H. L\"u, C.N. Pope, A. Sadrzadeh and T.A. Tran,
{\it Consistent $SO(6)$ reduction of type IIB supergravity on $S^5$},
Nucl. Phys. {\bf B586}, 275 (2000), hep-th/0003103.

\bm{con1} H. L\"u and C.N. Pope, 
{\it Exact embedding of $N = 1$, $D = 7$ gauged supergravity in $D = 11$}, 
Phys. Lett. {\bf B467}, 67 (1999), hep-th/9906168.

\bm{con2} M. Cveti\v c, H. L\"u and C.N. Pope,
{\it Gauged six-dimensional supergravity from massive type IIA}, 
Phys. Rev. Lett. {\bf 83}, 5226 (1999), hep-th/9906221.

\bm{con3}H. L\"u, C.N. Pope and T.A. Tran,
{\it Five-dimensional $N = 4$, $SU(2) \times U(1)$ gauged supergravity 
from type IIB},
Phys. Lett. {\bf B475}, 261 (2000), hep-th/9909203.

\bm{con4} M. Cveti\v c, H. L\"u and C.N. Pope,
{\it Four-dimensional $N = 4$, $SO(4)$ gauged supergravity from D = 11},
Nucl.\ Phys.\ B {\bf 574}, 761 (2000), hep-th/9910252.

\bm{spherered}
M. Cveti\v c, H. L\"u and C.N. Pope,
{\it Consistent Kaluza-Klein sphere reductions},
Phys. Rev. {\bf D62}, 064028 (2000), hep-th/0003286.

\bm{het2}
M.J. Duff, B.E.W. Nilsson, C.N. Pope and N.P. Warner,
{\it Kaluza-Klein approach to the heterotic string. 2},
Phys. Lett. {\bf B171} (1986) 170.

\bm{callan}
C.G. Callan, E.J. Martinec, M.J. Perry and D. Friedan,
{\it Strings in background fields},
Nucl. Phys. {\bf B262}, 593 (1985).

\bm{salsez8}A. Salam and E. Sezgin,
{\it $D = 8$ supergravity}, 
Nucl. Phys. {\bf B258}, 284 (1985).

\bibitem{cgr}
M. Cveti\v c, S. Griffies and S.J. Rey,
{\it Static domain walls in N=1 supergravity,}
Nucl.\ Phys.\ {\bf B381}, 301 (1992), hep-th/9201007.

\bibitem{cs}
M. Cveti\v c and H.H. Soleng,
{\it Supergravity domain walls,}
Phys.\ Rept.\  {\bf 282}, 159 (1997), hep-th/9604090.

\bm{dunipowa} M.J. Duff, B.E.W. Nilsson, C.N. Pope and N.P. Warner,
{\it On the consistency of the Kaluza-Klein ansatz}, 
Phys. Lett. {\bf B149}, 90 (1984).

\bm{homapo} P. Hoxha, R.R. Martinez-Acosta and C.N. Pope,
{\it Kaluza-Klein consistency, Killing vectors, and K\"ahler spaces}, 
Class. Quant. Grav. {\bf 17}, 4207 (2000),hep-th/0005172.


\end{thebibliography}
\end{document}